\documentclass[preprint,journal]{vgtc}            % arXiv preprint (journal style)
\usepackage[colorinlistoftodos]{todonotes}
\usepackage{listings}
\usepackage{tabularx}
\usepackage{multirow}

\definecolor{svgbg}{HTML}{F6F8FA}     % 极浅的灰底色
\definecolor{svgtag}{HTML}{22863A}    % 绿色用于标签名
\definecolor{svgstr}{HTML}{032F62}    % 深蓝色用于属性和字符串
\definecolor{svgcomment}{HTML}{6A737D} % 灰色用于注释/节点标记

\usepackage[most]{tcolorbox}
\tcbuselibrary{listings, skins}

\lstdefinestyle{prettysvg}{
    basicstyle=\ttfamily\scriptsize\linespread{1.1}, 
    moredelim=[s][\color{blue!70!black}]{<}{>},         
    moredelim=[s][\color{gray!70!black}\itshape]{[}{]}, 
    morestring=[b]",
    stringstyle=\color{red!60!black},                   
    showstringspaces=false,
    breaklines=false,                                   
    literate={×}{{$\times$}}1                           
}

\newtcblisting{svgbox}{
    enhanced,
    colback=gray!4,       
    colframe=gray!30,     
    boxrule=0.5pt,        
    arc=3pt,              
    left=4pt, right=4pt, top=4pt, bottom=4pt, 
    drop fuzzy shadow,  
    listing only,
    listing options={style=prettysvg}
}
\DeclareMathAlphabet{\mathcal}{OMS}{cmsy}{m}{n}

\newcommand{\etal}{\textit{et al.}\xspace}

\usepackage{tikz}

\definecolor{ContentFill}{HTML}{DCE8F8}
\definecolor{ContentBorder}{HTML}{8FB1E3}

\definecolor{ChartFill}{HTML}{F6D6D1}
\definecolor{ChartBorder}{HTML}{D88D82}

\definecolor{StyleFill}{HTML}{DCEECD}
\definecolor{StyleBorder}{HTML}{9AC57D}

\definecolor{LayoutFill}{HTML}{F7E8B5}
\definecolor{LayoutBorder}{HTML}{D6B656}

\definecolor{IllustrationFill}{HTML}{DDD0EE}
\definecolor{IllustrationBorder}{HTML}{B59ACF}

\definecolor{TaxHead}{HTML}{F4F1EA}

\definecolor{ContentRow}{HTML}{F5F9FE}
\definecolor{ContentCell}{HTML}{DCE8F8}
\definecolor{ContentBorder}{HTML}{8FB1E3}

\definecolor{ChartRow}{HTML}{FDF4F2}
\definecolor{ChartCell}{HTML}{F6D6D1}
\definecolor{ChartBorder}{HTML}{D88D82}

\definecolor{LayoutRow}{HTML}{FDF9EC}
\definecolor{LayoutCell}{HTML}{F7E8B5}
\definecolor{LayoutBorder}{HTML}{D6B656}

\definecolor{IllustrationRow}{HTML}{F8F5FC}
\definecolor{IllustrationCell}{HTML}{DDD0EE}
\definecolor{IllustrationBorder}{HTML}{B59ACF}

\definecolor{StyleRow}{HTML}{F6FBF2}
\definecolor{StyleCell}{HTML}{DCEECD}
\definecolor{StyleBorder}{HTML}{9AC57D}

\onlineid{0}

\preprinttext{arXiv preprint}

\vgtccategory{Research}

\vgtcpapertype{system}

\title{Show Me the Infographic I Imagine: Intent-Aware Infographic Retrieval for Authoring Support}

\author{%
  Jing Xu,
  Jiarui Hu,
  Zhihao Shuai,
  Yiyun Chen, and
  Weikai Yang
}

\authorfooter{
  \item Jing Xu, Zhihao Shuai, Yiyun Chen, and Weikai Yang are with The Hong Kong University of Science and Technology (Guangzhou).
  \item Jiarui Hu is with Nanjing University.
  \item Corresponding author: Weikai Yang.
}

\abstract{%
While infographics have become a powerful medium for communicating data-driven stories, authoring them from scratch remains challenging, especially for novice users. 
Retrieving relevant exemplars from a large corpus can provide design inspiration and promote reuse, substantially lowering the barrier to infographic authoring.
However, effective retrieval is difficult because users often express design intent in ambiguous natural language, while infographics embody rich and multi-faceted visual designs.
As a result, keyword-based search often fails to capture design intent, and general-purpose vision-language retrieval models trained on natural images are ill-suited to the text-heavy, multi-component nature of infographics.
To address these challenges, we develop an intent-aware infographic retrieval framework that better aligns user queries with infographic designs.
We first conduct a formative study of how people describe infographics and derive an intent taxonomy spanning content and visual design facets. 
This taxonomy is then leveraged to enrich and refine free-form user queries, guiding the retrieval process with intent-specific cues.
Building on the retrieved exemplars, users can adapt the designs to their own data with high-level edit intents, supported by an interactive agent that performs low-level adaptation.
Both quantitative evaluations and user studies are conducted to demonstrate that our method improves retrieval quality over baseline methods while better supporting intent satisfaction and efficient infographic authoring.

}

\keywords{Infographic retrieval, infographic authoring, design intent, interactive design adaptation}

\teaser{
  \centering
  \includegraphics[width=\linewidth, alt={A view of a city with buildings peeking out of the clouds.}]{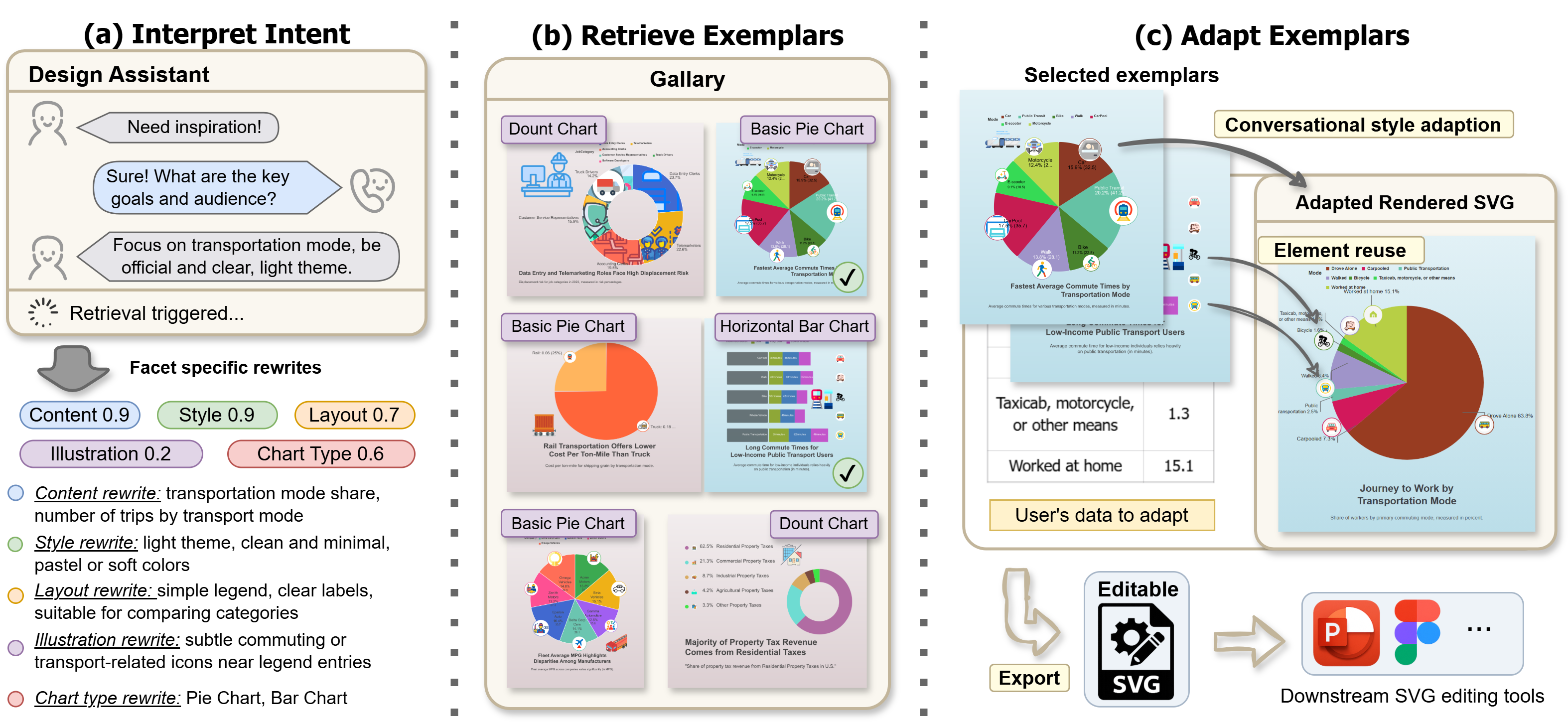}
  \caption{
    Overview of our intent-aware retrieve-and-adapt workflow for infographic authoring.
  (a) A natural-language query is parsed into weighted intent facets;
  (b) These facet-specific cues guide exemplar retrieval over an infographic corpus;
  (c) The selected exemplar is adapted to the user's data, producing a rendered result together with an editable SVG that can be exported to downstream design tools.
  }
  \label{fig:teaser}
}

\renewcommand{\manuscriptnotetxt}{}

\graphicspath{{figs/}{figures/}{pictures/}{images/}{./}} % where to search for the images

\usepackage{booktabs}                  % only used for the table example
\usepackage{lipsum}                    % used to generate placeholder text
\usepackage{mwe} 
\usepackage{amsmath} 
\usepackage{amssymb} % for \mathbb
\usepackage{mathptmx}                  % use matching math font
\usepackage[table]{xcolor}
\usepackage{array}
\usepackage{booktabs}
\usepackage{ragged2e}

\definecolor{taxhead}{HTML}{D8C6C6}
\definecolor{taxfacet}{HTML}{E8D6C7}
\definecolor{taxcap}{HTML}{C9D9D0}
\definecolor{taxcue}{HTML}{DDD6E8}

\newcolumntype{L}[1]{>{\RaggedRight\arraybackslash}p{#1}}
\newcolumntype{Y}{>{\RaggedRight\arraybackslash}X}
\begin{document}

%%%%%%%%%%%%%%%%%%%%%%%%%%%%%%%%%%%%%%%%%%%%%%%%%%%%%%%%%%%%%%%%
%%%%%%%%%%%%%%%%%%%%%% START OF THE PAPER %%%%%%%%%%%%%%%%%%%%%%
%%%%%%%%%%%%%%%%%%%%%%%%%%%%%%%%%%%%%%%%%%%%%%%%%%%%%%%%%%%%%%%%

\firstsection{Introduction}

\maketitle

Infographics are a widely used medium for communicating data-driven stories by combining data, text, and visual elements into a compact narrative~\cite{wang2022mixedinitiative_reuse, elaldi2021effectiveness,Liu2025}.
Yet authoring a high-quality infographic remains challenging, as authors must coordinate multiple design decisions simultaneously, including narrative flow, layout composition, visual style, and illustration usage.
In practice, both novice users and expert designers often start from existing designs by reusing templates or borrowing design patterns to reduce effort and improve quality~\cite{lee2010designing}.
This motivates exemplar retrieval as a practical authoring aid: given a user's design intent, relevant exemplars that match not only the topic but also design intent should be retrieved.

However, existing retrieval methods often fail to satisfy users' design intent for infographics.
Traditional keyword-based search engines~\cite{saleh2015learning, 9903579} primarily optimize for topic matching and provide limited support for expressing and enforcing structural or stylistic constraints.
Meanwhile, embedding-based text--image retrieval models~\cite{radford2021clip, jia2021align} typically collapse relevance into a single similarity, which offers limited control over which intent facet (e.g., layout composition or visual style) should dominate the retrieval.
As a result, models frequently return exemplars that are semantically related but misaligned with the desired layout or style.
These limitations motivate an intent-aware retrieval method that can better capture and enforce multiple design intent facets.
% For example, a user may ask for a step-by-step process infographic with minimal icons and muted pastel tones, yet conventional retrieval can return process-related results that ignore the step layout or stylistic constraints.

To characterize this mismatch, we first conducted a formative study to examine how people describe desired infographic exemplars in free-form natural language.
Our analysis showed that queries commonly blend multiple intent facets, including content, chart type, layout, illustration, and style.
%  \contentfacet{content}, \chartfacet{chart type}, \layoutfacet{layout}, \illustrationfacet{illustration}, and \stylefacet{style}.
Guided by these findings, we propose an intent-aware retrieval framework that represents a user's requirement as a weighted combination of these intent facets.
Given a free-form query, the system generates facet-specific rewrites, estimates facet weights, and performs weighted multi-facet matching in a shared text--image embedding space.
To better support infographic retrieval, we additionally apply a lightweight embedding alignment technique to better capture facet-specific query--infographic similarity.

Beyond retrieval, we present an exemplar-driven authoring workflow that helps users reuse and adapt retrieved designs to their own data.
The workflow is supported by an interactive conversational agent that translates high-level edit intents into concrete modifications of the exemplar representation (e.g., SVG-based edits), thereby connecting inspiration seeking with iterative adaptation~\cite{yang2024foundation}.

% Original: We evaluate our method through both single-round retrieval benchmarks and a multi-round, authoring-oriented user study.
We evaluate our method through single-round and multi-round retrieval benchmarks, as well as an end-to-end authoring user study.
The results indicate that our method improves retrieval quality over baseline techniques and better supports intent satisfaction during authoring.
The contributions of our work are as follows:
\begin{itemize}[nosep]
  % Original: We conducted a formative study of how people search for infographics and derive a structured intent taxonomy spanning five facets.
  \item We conducted a formative study of how people describe desired infographic exemplars in free-form natural language and derived a structured intent taxonomy spanning five facets.
  \item We introduce an intent-aware infographic retrieval framework that leverages taxonomy-guided query rewriting, facet weighting, and lightweight embedding alignment for multi-facet matching.
  \item We build an interactive system that supports end-to-end exemplar-driven authoring in a chat session, enabling users to retrieve relevant exemplars and adapt them to their own data.
\end{itemize}
\section{Related Work}

\subsection{Infographic Authoring}
Research on infographic authoring can be divided into two lines of work: authoring from scratch\cite{liu2018dataillustrator, ren2019charticulator, wang2018infonice, wang2023nlvisauthoring, wang2024dataformulator} and authoring through exemplar reuse~\cite{qian2021retrieve_then_adapt, wang2022mixedinitiative_reuse}.
In the first line, expressive authoring tools such as Data Illustrator~\cite{liu2018dataillustrator} and Charticulator~\cite{ren2019charticulator} give authors direct control over marks, bindings, and layout constraints, while systems like InfoNice~\cite{wang2018infonice} package similar capabilities into more novice-friendly workflows.
Recent mixed-initiative systems further lower the barrier by interpreting higher-level user goals expressed in natural language~\cite{wang2023nlvisauthoring,wang2024dataformulator} or structured around the message to be conveyed~\cite{zhou2024epigraphics}.
However, these tools still require numerous design decisions, which can be overwhelming for authors without a clear design goal.
Since designers frequently start from existing designs to reduce effort~\cite{lee2010designing}, a second line of work supports authoring through exemplar reuse.
Retrieve-Then-Adapt~\cite{qian2021retrieve_then_adapt} retrieves a proportion-related infographic exemplar and adapts it to new data, and Wang \etal~\cite{wang2022mixedinitiative_reuse} generalized this idea by treating existing infographic charts as reusable authoring assets.
Complementary systems automate narrower reuse subproblems such as timeline structure extraction~\cite{chen2020automated_infographic_timeline}, design exploration~\cite{tyagi2022infographics_wizard}, and palette recommendation~\cite{yuan2022infocolorizer}.
In contrast to these methods, our method focuses on the missing retrieval layer between vague author intent and downstream adaptation.

\subsection{Visualization Recommendation}
Visualization recommendation automatically generates and ranks chart specifications from structured data.
Research in this area has progressed from rule- and specification-based chart suggestion~\cite{mackinlay2007showme,wongsuphasawat2016voyager,wongsuphasawat2016compassql,wongsuphasawat2017voyager2,satyanarayan2017vegalite}, to knowledge-based recommenders~\cite{moritz2019draco,hu2019vizml,li2022kg4vis}, and more recent natural-language, user-adaptive systems~\cite{narechania2021nl4dv,data2vis,song2022rgvisnet,zhang2024adavis,guo2025more_like_vis,davis2024risks_of_ranking}. % retrieval-augmented,
Early systems such as \emph{Show Me}~\cite{mackinlay2007showme} and \emph{Voyager}~\cite{wongsuphasawat2016voyager} frame design rules as search over a structured space of encodings and transformations, while grammars such as Vega-Lite~\cite{satyanarayan2017vegalite} make it easier to steer recommendation from partial specifications.
While this line of work makes design choices easier, it still depends heavily on hand-authored grammars and explicitly specified constraints.
Later work responds by learning or formalizing richer ranking criteria.
For example, \emph{VizML}~\cite{hu2019vizml} learns likely design choices from large corpora, while \emph{KG4Vis}~\cite{li2022kg4vis} makes these criteria more explicit and interpretable using knowledge graphs.
Recent systems make user intent easier to express, either by mapping natural-language inputs to chart specifications~\cite{narechania2021nl4dv,data2vis,shuai2026cluster} or by incorporating preference signals~\cite{song2022rgvisnet,zhang2024adavis,guo2025more_like_vis,davis2024risks_of_ranking}.
These advances are informative for our setting, but recommendation systems typically output chart specifications over structured data, whereas we retrieve whole infographic references whose usefulness depends on holistic design factors that such systems usually abstract away.

% Original: Recent systems then make user intent easier to express and refine, either by mapping utterances to chart specifications or programs, as in \emph{NL4DV}~\cite{narechania2021nl4dv} and \emph{Data2Vis}~\cite{data2vis}, or by incorporating retrieval and preference signals, as in \emph{RGVisNet}~\cite{song2022rgvisnet} and preference-aware recommendation interfaces~\cite{zhang2024adavis,guo2025more_like_vis,davis2024risks_of_ranking}.

\subsection{Visualization Retrieval and Cross-Modal Search}
Unlike recommendation that constructs specifications within a predefined grammar space, retrieval searches a corpus of existing visualizations to find relevant examples.
This is particularly suited for infographic authoring, where users often seek design references that are difficult to specify in advance.
Cross-modal visual retrieval in this context includes three related lines of work: generic vision-language retrieval models~\cite{radford2021clip,jia2021align,tschannen2025siglip,zhou2025megapairs}, example-centric and context-aware design search~\cite{lee2010designing,9903579,kovacs2019contextawareassetsearch,son2024genquery}, and visualization-specific retrieval methods~\cite{saleh2015learning,oppermann2021vizcommender,li2022structureaware,nowak2024multimodalchartretrieval,chen2025chart2vec,nguyen2025safire}.
Generic models such as \emph{CLIP}~\cite{radford2021clip} align text and images in a shared embedding space to enable open-ended retrieval, and later large-scale training efforts such as \emph{SigLIP}~\cite{tschannen2025siglip} and \emph{MegaPairs}~\cite{zhou2025megapairs} further strengthen this paradigm.
However, generic embedding similarity typically collapses multiple relevance cues into a single score, offering limited transparency and limited control over which aspect of a design dominates matching.
A complementary line of work in design studies and design-support tools highlights the central role of examples in creative practice~\cite{lee2010designing,9903579}.
Building on this insight, Kovacs \etal~\cite{kovacs2019contextawareassetsearch} ranked candidate assets based on their compatibility with the surrounding composition, while Son \etal~\cite{son2024genquery} supported more expressive search by helping users concretize vague intents and search with generated or edited visual queries.
In the visualization domain, Saleh \etal~\cite{saleh2015learning} modeled style similarity for infographic search, capturing aesthetic resemblance but not broader semantic or structural intent.
Later work incorporates richer retrieval signals, such as structural and visual cues~\cite{li2022structureaware}, comparisons of chart retrieval pipelines~\cite{nowak2024multimodalchartretrieval}, and explicit consideration of which visualization properties should determine similarity~\cite{nguyen2025safire}.
Our work builds on these works but focuses on infographic exemplar search for authoring, where the query is often under-specified and must be decomposed into controllable intent facets rather than collapsed into a single similarity notion.

\section{Formative Study}
\label{sec:formative}
\begin{figure}
    \centering
    \includegraphics[width=\linewidth]{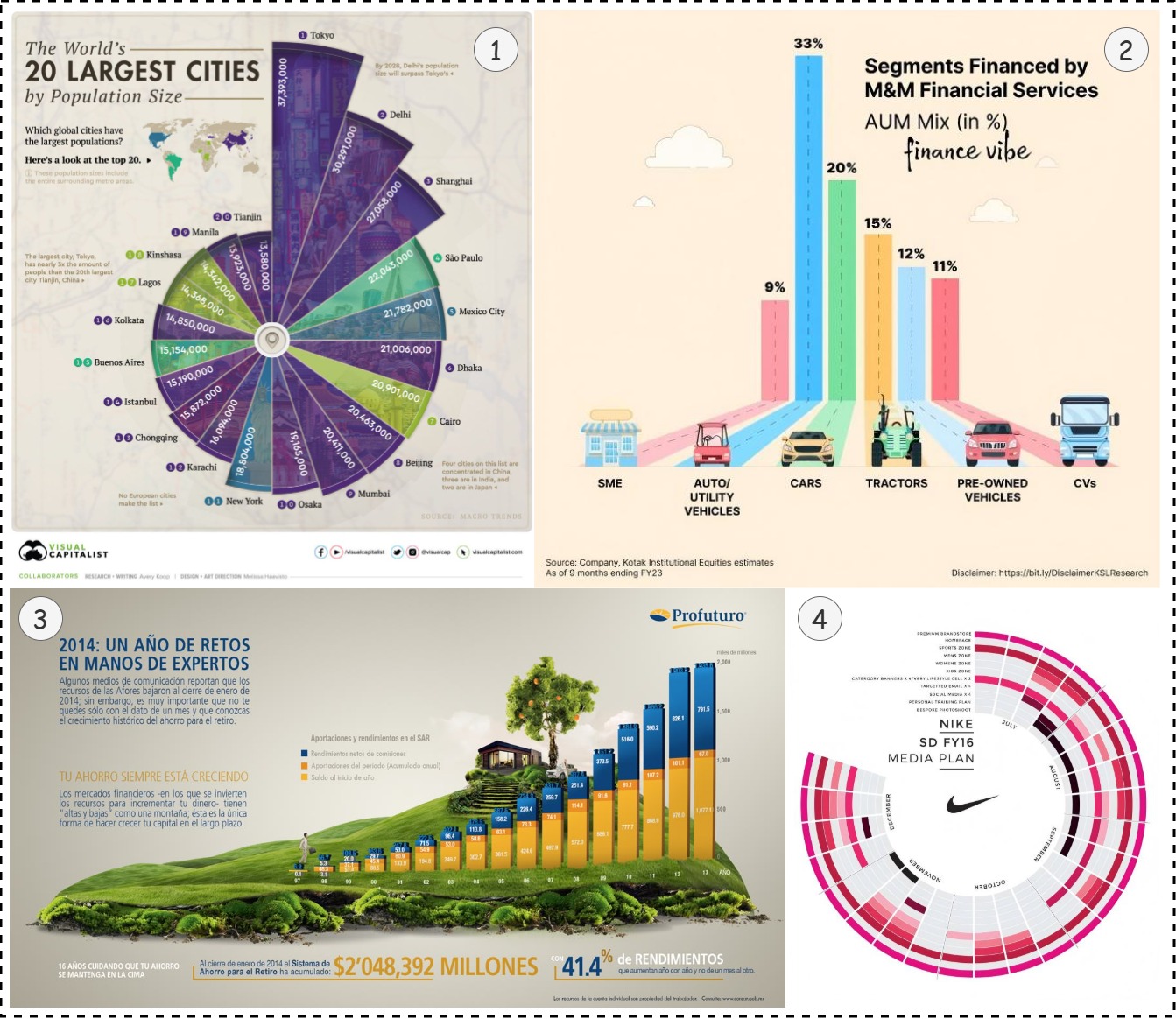}
    \caption{Four infographic exemplars used in the formative study to elicit participants' queries of desired infographic exemplars.}
    \label{fig:exemplars}
\end{figure}

\begin{table*}[t]
\centering
\small
\caption{\textbf{Five-facet intent taxonomy derived from the formative study.}
The taxonomy summarizes recurring facets in users' natural-language descriptions of desired infographic exemplars.
Facets can be partially specified: a query may omit one or more facets, or express them only implicitly.}
\label{tab:intent-taxonomy}

\renewcommand{\arraystretch}{1.18}
\setlength{\tabcolsep}{6pt}

% Original: \begin{tabularx}{\textwidth}{L{0.14\textwidth}YY}
\begin{tabularx}{\textwidth}{>{\RaggedRight\arraybackslash}m{0.14\textwidth}YY}
\toprule
\rowcolor{TaxHead}
\textbf{Facet} & \textbf{What it captures} & \textbf{Typical cues in user queries} \\
\midrule

\rowcolor{ContentRow}
\cellcolor{ContentCell}\textbf{Content} &
Communicative goal and information organization: what should be shown and what takeaway should be emphasized. &
category breakdown/share; trend or growth over time; compare A vs.\ B; highlight the largest component; include key numbers or summaries \\

\rowcolor{StyleRow}
\cellcolor{StyleCell}\textbf{Style} &
Holistic visual aesthetics: overall tone and look-and-feel, including palette and typography cues. &
clean or minimalist; editorial or magazine-like; warm or playful; brand-like; pastel or muted colors; 3D or colorful; typography-forward \\

\rowcolor{LayoutRow}
\cellcolor{LayoutCell}\textbf{Layout} &
Spatial composition and narrative organization: how the infographic is arranged and read, beyond chart type alone. &
vertical poster; clear hierarchy or sections; center chart with side summaries; radial layout with fixed-angle segments; dense labels or annotations; start at the positive $y$-axis and go clockwise \\

\rowcolor{IllustrationRow}
\cellcolor{IllustrationCell}\textbf{Illustration} &
Whether and how icons or illustrations are used, including their density and explanatory versus decorative role. &
icons for each category; illustration-heavy; replace labels with pictorial symbols; scene-based background; minimal decoration \\

\rowcolor{ChartRow}
\cellcolor{ChartCell}\textbf{Chart Type} &
Primary chart or visual form used to encode data, often as shorthand for the intended visual encoding. &
bar chart; stacked bars; pie or donut; rose/Nightingale chart; radial chart; ring heatmap; timeline-like chart \\

\bottomrule
\end{tabularx}
\end{table*}

\subsection{Study Goals and Research Questions}
We conducted a formative study to inform the design of our intent-aware infographic retrieval system by understanding how users describe their search intent.
Specifically, we elicited participants' queries of desired exemplars under two interfaces: a conventional keyword-based image search interface, and an imagined AI-assisted retriever that can interpret long natural language prompts.
This study design allows us to characterize the facets of intent that users naturally include and identify gaps in expression when users are constrained to keyword-based search.

Accordingly, we formulated two research questions:
\begin{itemize}[nosep]

  \item \textbf{RQ1:} What facets of intent do users usually express in natural language descriptions of desired infographics?
  \item \textbf{RQ2:} Which facets are under-expressed in keyword queries compared to natural language queries, and how does this affect the retrieval performance and users' confidence?
\end{itemize}

Natural language queries are hypothesized to capture richer intent context than keyword queries.
By answering RQ1 and RQ2, we seek to validate this hypothesis and identify specific intent facets that a keyword-based interface might fail to capture.
These insights will directly inform the design of the query interpretation module in our system and highlight where intent-aware features are most needed.

\subsection{Participants}
% Original: We recruited \textbf{14} participants to complete an online questionnaire.
We recruited \textbf{14} participants (\(n{=}14\); \(10\) male / \(4\) female), who completed the online questionnaires remotely.
% --- new ---
Participants were aged 21--45 years (\textit{M}=27.1, \textit{SD}=6.8) and reported majors/primary fields in computer science and design.
Most participants were students (1 undergraduate / 7 master's / 3 PhD), and the remainder were practitioners working in data analysis or software engineering.
% ---
% [done] xujing, todo: add more details about the participants, e.g. age, major, etc.
% Original: Fourteen participants completed the online questionnaire (\(n{=}14\); \(10\) male / \(4\) female).
% Fourteen participants completed the online questionnaire (\(n{=}14\); \(10\) male / \(4\) female).
% Original: Participants were recruited online and completed the study remotely.
% Participants completed the questionnaire remotely.

To contextualize participants' familiarity with data visualization, we asked them to self report on 1) how often they create charts/visualizations and 2) how often they search online for reference charts/infographics for design inspiration.
For chart creation, some (\(5/14\)) reported doing so frequently, and most (\(9/14\)) reported doing so occasionally.
For searching reference charts, most (\(8/14\)) reported doing so frequently, some (\(5/14\)) reported doing so occasionally, and only one (\(1/14\)) reported doing so almost never.
This indicates that our participants have a wide range of prior exposure to chart-making and design-inspiration seeking behaviors, which is relevant for interpreting differences in how participants articulate infographic intent.
% can reduce if necessary

\begin{figure*}[t]
  \centering
  \includegraphics[width=\textwidth]{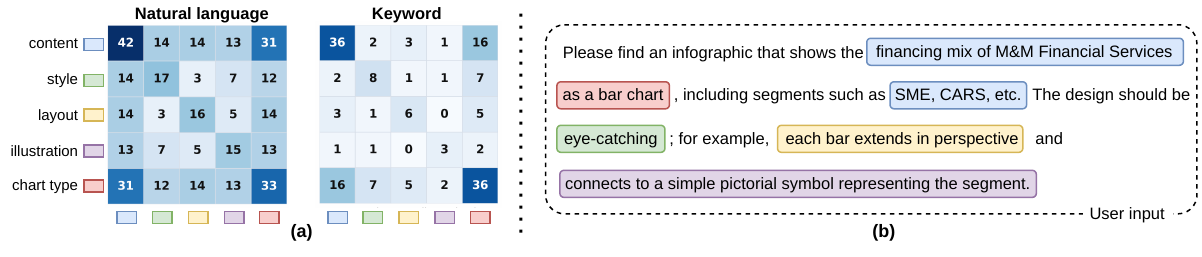}
  \caption{
  \textbf{Facet patterns in participants' queries.}
  (a) A facet co-occurrence matrix computed from the natural language queries and keyword queries.
  (b) An annotated example query illustrates how a single query can contain multiple facet signals, motivating facet-specific rewriting in retrieval.
  }
  \label{fig:formative-facets}
  \end{figure*}
\subsection{Procedure}
% Original: To encourage participants to describe visual and structural properties beyond topical keywords, we selected four infographic exemplars that differ in chart family, layout structure, illustration density, and overall aesthetic, including:
To cover a diverse range of infographic designs for eliciting query descriptions, we selected four infographic exemplars that differ in chart type, layout, illustration density, and overall aesthetic. As shown in Figure \ref{fig:exemplars}, they include:
1) a minimalist radial schedule / calendar-like visualization, %  (concentric rings with monthly segments)
2) an illustrated, perspective bar-chart poster, %  (categories visualized as bars emerging from a stylized road)
3) a scene-based, 3D-styled growth infographic, %  (time-series bars embedded in a natural landscape)
and 4) an editorial-style radial ranking infographic with dense annotations.

For each stimulus image, participants completed the following tasks:
\begin{enumerate}[nosep]
  \item \textbf{Keyword-style query.}
  % Original: Participants imagined searching on a typical image search website (e.g., Pinterest or Google Images) and wrote what they would realistically type into the search box based on their existing habits.
  Participants imagined using a typical image search website (e.g., Pinterest or Google Images) and wrote the query they would realistically type based on their usual search habits.
  % Original: Participants also rated how likely their keyword query would retrieve their desired results based on their experiences on image search websites a 5-point Likert scale (\(1=\) very unlikely, \(5=\) very likely)..
  They then rated, on a 5-point Likert scale (\(1=\) very unlikely, \(5=\) very likely), how likely this keyword query would be to retrieve their desired results based on their prior experience with such websites.
  \item \textbf{Natural language query.}
  % Original: Participants imagined a more AI-powered infographic retrieval system that could understand long, prompt-like natural language descriptions.
  Participants imagined an AI-powered infographic retrieval system that could understand long, prompt-like natural language descriptions.
  % Original: Ignoring current technical limitations, they wrote what they \emph{wish} they could input to retrieve the target infographic or highly similar ones.
  Ignoring current technical limitations, they wrote what they would ideally like to input to retrieve the target infographic or highly similar ones.
  % Original: Note that we do not collect confidence ratings for the natural language query, because users do not have experiences on NL-query-based infographic retrieval.
  We did not collect confidence ratings for the natural language query because participants generally lacked prior experience with such systems.

\end{enumerate}
\vspace{0.3em}
We randomized the order of the four stimuli to mitigate potential order effects.
The study yielded 56 (\(4 \times 14\)) keyword-style queries with confidence ratings and 56 natural language queries.

\subsection{Results and Analysis}
We conducted iterative qualitative coding to identify recurring intent facets in both keyword queries and natural language queries.
Two researchers independently performed open coding on a subset of queries, marking phrases that describe \emph{what} the infographic should communicate and \emph{how} it should be visually realized.
The two researchers then reconciled differences through discussion, consolidated an initial codebook, and iteratively refined it while applying the codes to the full set of queries.
Here, multiple codes per query are allowed because participants frequently combined multiple requirements in a single query.
Finally, the two researchers grouped codes into higher-level categories that capture major intent facets and examined how these facets co-occur within individual queries.
This coding process yields three findings that directly motivate an intent-aware retrieval formulation.

\paragraph{F1: Users describe infographics through multiple interacting facets.}
Participants often went beyond topic keywords and described design-relevant aspects, such as information organization, compositional structure, illustration usage, and overall aesthetic.
For example, one participant described the second infographic as ``each bar extends in perspective,'' which captures both the chart type and the stylistic rendering rather than only the topical content.
Participants also used chart-type terms as an explicit part of their intent (e.g., ``rose chart,'' ``ring heatmap,'' ``bar chart'') rather than as an afterthought.
From these descriptions, we derived a five-facet taxonomy of infographic search intent:
\textbf{content} (communicative goal and what should be conveyed),
\textbf{chart type} (the primary chart/visual form used to encode data),
\textbf{layout} (spatial composition, hierarchy, and narrative organization),
\textbf{illustration} (icons/illustrations and their usage),
and \textbf{style} (overall aesthetics such as typography, palette, and visual tone).
Table~\ref{tab:intent-taxonomy} summarizes the facets and typical linguistic cues observed in the study.

We further observe that facets often co-occur within a single query, and that a query can mix explicit constraints (e.g., ``radial layout'') with implicit signals (e.g., ``editorial-style'' implying dense annotations and typography-heavy design).
% Original: Figure~\ref{fig:formative-facets} visualizes these patterns via a facet co-occurrence matrix and an annotated example query, motivating our facet-specific rewriting and weighting strategy for retrieval.
Figure~\ref{fig:formative-facets} shows a facet co-occurrence matrix and an annotated example query.

\paragraph{F2: Natural language queries encode richer multi-facet intent, whereas keyword queries compress intent and under-specify design constraints.}

% Original: Across the two interface conditions (\(n{=}56\) AI-style; \(n{=}56\) traditional keyword), AI-style inputs consistently contained more facet-level constraints and were more likely to include design-critical requirements beyond topical semantics.
% Across the two interface conditions (\(n{=}56\) natural language; \(n{=}56\) keyword), natural language inputs contained more facet-level constraints and more often included design-critical requirements beyond topical semantics.
When comparing natural language queries and keyword queries, natural language ones expressed more facet-level constraints and more often included design-critical requirements beyond topical semantics.
Specifically, natural language queries expressed more facets per query (\(2.29\) facets/query) than keyword queries (\(1.59\) facets/query).
% Original: Moreover, half of the AI-style prompts (\(28/56=50.0\%\)) contained \(\ge 3\) facets, while only \(5/56=8.9\%\) of traditional keyword queries reached this level, indicating that keyword inputs often compress intent into one or two high-level descriptors.
Moreover, half of the natural language queries contained \(\ge 3\) facets, while only \(8.9\%\) of keyword queries reached this level, indicating that keyword inputs often compress intent into one or two high-level descriptors.
% Original: This compression disproportionately affected design-critical facets.
A detailed analysis reveals that this compression disproportionately affected design-critical facets.
Natural language queries frequently specified \emph{layout} (\(39.3\%\)), \emph{illustration usage} (\(28.6\%\)), and \emph{style/aesthetic tone} (\(32.1\%\)), whereas these facets were rare in keyword queries (layout \(10.7\%\), illustration \(5.4\%\), and style \(14.3\%\)). 
Aggregating these three facets, \(69.6\%\) of natural language queries mentioned at least one design facet (layout, illustration, or style), compared to \(26.8\%\) of keyword queries. 
In contrast, content and chart-type terms remained common in both conditions (\(71.4\%\) and \(64.3\%\) for natural language and keyword queries, respectively).

% We treat these as ``no-facet'' queries in the quantitative summaries above.
% In the qualitative review, they suggest that when users imagine a more AI-powered infographic retrieval system, they may also expect multimodal ways to convey intent beyond textual facet descriptions.
% Together, these results suggest that natural language queries elicit richer and more explicit intent---especially for layout, illustration, and style---whereas keyword queries tend to under-express these design constraints.

\paragraph{F3: Users are not confident that keyword search will return their intended designs.}
Participants' confidence ratings provide quantitative evidence for this gap.
Across 56 keyword-based queries, the median confidence was only \(2\) (IQR: \(1\)–\(3\); \(M{=}2.32\), \(SD{=}1.15\)).
Moreover, \(82.1\%\) of the ratings were \(3\) or below, suggesting that participants generally did not expect conventional keyword-based search to reliably return their intended results even when a clear target was shown.
This aligns with a common failure mode in practice: results can be semantically related yet misaligned with design-critical facets such as chart type, layout, illustration usage, and overall style.

Together, these findings motivate an intent-aware retrieval approach that 1) makes intent facets explicit, 2) supports partial specification when some facets are absent, and 3) performs facet-aware matching rather than relying on a single overall similarity score.

\section{Intent-Aware Infographic Retrieval Framework}
\label{sec:method}

\begin{figure*}[t]
  \centering
  \includegraphics[width=\linewidth]{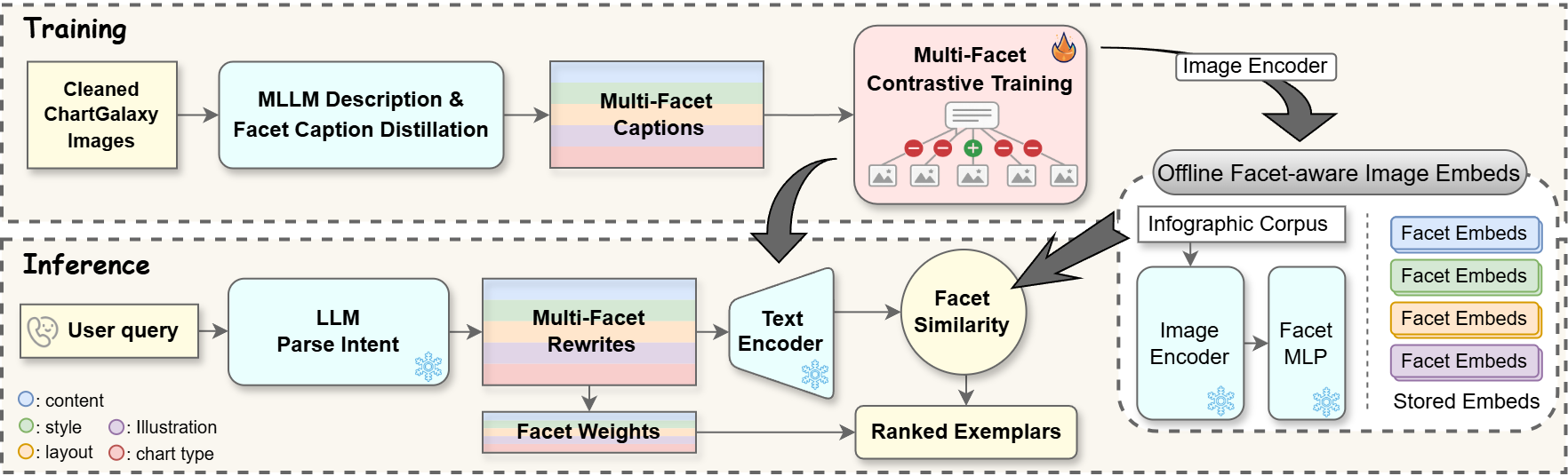}
  \caption{\textbf{Overview of our intent-aware infographic retrieval framework.}
  \emph{Top (training):} ChartGalaxy images are described by a multimodal LLM and distilled into short facet-specific captions (two per facet). We fine-tune the full model (text encoder, image encoder and facet MLP heads) with a multi-facet in-batch contrastive loss (Eq.~\ref{eq:facet-infonce}).
  \emph{Bottom (inference):} A user query $q$ is parsed into facet rewrites $\{q_f\}$, facet weights $\mathbf{w}$, and an optional multi-choice chart-type set $\mathcal{T}_q$. We compute facet-conditioned text embeddings and compare them with precomputed facet-specific image embeddings to obtain $\{s_f\}_{f\in\mathcal{F}_e}$, and compute a multi-choice discrete chart-type similarity $s_{\textsc{T}}$. 
  We fuse all facets by a weighted sum to rank exemplars.}
  \label{fig:framework}
\end{figure*}

Motivated by the multi-faceted nature of infographic search intent, we propose an intent-aware retrieval framework that 1) parses a free-form query into five intent facets with per-facet weights, 2) computes facet-aware similarities, and 3) ranks exemplars based on a weighted combination of facet-aware similarities.
Figure~\ref{fig:framework} summarizes the pipeline.

\subsection{Intent Representation and Query Parsing}
% Given an infographic corpus $\mathcal{X}=\{x_i\}_{i=1}^{N}$ and a free-form user query $q$, our goal is to retrieve a ranked list of exemplars that best match the user’s \emph{design intent}.
% Our formative study (Section~\ref{sec:formative}) shows that users describe desired exemplars through multiple facets (Table~\ref{tab:intent-taxonomy}), and that different facets may be present, absent, or emphasized to different degrees.
% We therefore model a query as a weighted combination of five intent facets:
% \textbf{content} (\textsc{C}), \textbf{chart type} (\textsc{T}), \textbf{layout/structure} (\textsc{L}), \textbf{illustration} (\textsc{I}), and \textbf{style} (\textsc{S}).
% Let $\mathcal{F}=\{\textsc{C},\textsc{T},\textsc{L},\textsc{I},\textsc{S}\}$ and $\mathcal{F}_e=\mathcal{F}\setminus\{\textsc{T}\}=\{\textsc{C},\textsc{L},\textsc{I},\textsc{S}\}$ denote the embedding-based facets.

Given a free-form query $q$, we parse it into five intent facets $\mathcal{F}=\{\textbf{content}~(\textsc{C}),~\textbf{chart type}~(\textsc{T}),~\textbf{layout}~(\textsc{L}),~\textbf{illustration}~(\textsc{I}),~\textbf{style}~(\textsc{S})\}$.
Among these, chart type takes values from a relatively fixed label space $\mathcal{T}$, so we treat it as a multi-choice constraint and infer a set of labels $q_{\textsc{T}}\subseteq\mathcal{T}$ from $q$ when specified.
Following ChartGalaxy~\cite{li2026chartgalaxy}, we use a compact pool $\mathcal{T}$ of 13 coarse chart types:
\emph{Bar Chart, Line Chart, Area Chart, Radar Chart, Pie Chart, Scatterplot, Gauge Chart, Treemap, Diagram, Histogram, Range Chart, Funnel Chart, Pyramid Chart}.
The remaining four facets are naturally expressed in open vocabulary, so we use LLMs to rewrite $q$ into concise facet-focused descriptions $\{q_f\}_{f\in\{\textsc{C},\textsc{L},\textsc{I},\textsc{S}\}}$.
For those unspecified facets, we set $q_f=\emptyset$.

In addition to the facet-focused descriptions, the parser also predicts a non-negative facet weight vector $\mathbf{w}=\{w_f\}_{f\in\mathcal{F}}$ to reflect the importance of each facet.
We set $w_f=0$ if the facet $f$ is unspecified. % (i.e., $q_f=\emptyset$)

In our implementation, we use Qwen3-32B with a taxonomy-guided prompt and a fixed output schema to parse the query. We validate the schema and automatically retry on invalid outputs. The first-time schema invalid rate is about 4e-4.
Prompt templates and the full schema are provided in Supplementary Section~2.
% \vicacomment{[schema validity rate and average retries.]}
% and automatically validate and retry on invalid outputs; we validate the schema and automatically retry on invalid outputs.

% We denote the embedding facets as $\mathcal{F}_e=\{\textsc{C},\textsc{L},\textsc{I},\textsc{S}\}$.
% Following ChartGalaxy \cite{li2026chartgalaxy}, we abstract fine-grained types into a compact pool of 13 coarse types:
% \emph{Bar Chart, Line Chart, Area Chart, Radar Chart, Pie Chart, Scatterplot, Gauge Chart, Treemap, Diagram, Histogram, Range Chart, Funnel Chart, Pyramid Chart}.
% In contrast, the remaining facets are expressed in open vocabulary.
% Therefore, we use the same LLM to extract facet-relevant cues and rewrite them into concise, facet-focused texts for robust comparison in embedding space.
% Unspecified facets are set to $\emptyset$.
% In our implementation, we use Qwen3-32B with a taxonomy-guided prompt and a fixed output schema, as it provides reliable structured outputs while offering a practical balance between quality and computational cost.
% We validate the schema and automatically retry on invalid outputs.
% Prompt templates and the full schema are provided in the supplementary material.

\subsection{Facet-Aware Retrieval Scoring}
After parsing $q$ into facets, we score each exemplar infographic $x_i$ by combining multi-facet similarities.
% We fuse all facet-level similarities using the facet weights $\mathbf{w}$ predicted during parsing.
The final relevance score is a weighted sum $S(q,x_i)= \sum_{f\in\mathcal{F}} w_f \cdot s_f(q,x_i)$, where $s_f(q,x_i)$ is the similarity score for facet $f$ between the query $q$ and the exemplar $x_i$.
We next describe how we compute $s_f(q,x_i)$ for each facet.

\subsubsection{Chart Type Similarity}
In ChartGalaxy, each exemplar $x_i$ is associated with a chart type set $T_{x_i}\subseteq\mathcal{T}$.
When the user specifies chart type as a multi-choice set $q_{\textsc{T}}\subseteq\mathcal{T}$, we compute a chart-type similarity score $s_{\textsc{T}}(q,x_i)$ between $q_{\textsc{T}}$ and $T_{x_i}$.
Rather than requiring set-overlap agreement with strict matches, we allow \emph{soft} matches between visually similar chart types, since they can be ambiguous at coarse granularity and users may not always name the intended type precisely (e.g., an area chart can be described as a line chart with the region filled).

Concretely, we define a type-to-type kernel $\kappa(t,t')\in[0,1]$ with $\kappa(t,t')=1$ if $t=t'$, and non-zero values only for expert-specified similar pairs (listed in Supplementary Section~3).
We then match each queried type to the best-matching type present in the exemplar and average across queried types:
\begin{equation}
\label{eq:chart-sim}
s_{\textsc{T}}(q,x_i)=
\frac{1}{|\mathcal{T}_q|}\sum_{t\in \mathcal{T}_q} \max_{t'\in T_{x_i}} \kappa(t, t'),
\end{equation}
\subsubsection{Embedding-Based Facet Similarities.}
For the open-vocabulary facets $\{\textsc{C},\textsc{L},\textsc{I},\textsc{S}\}$, we compute a separate text--image similarity per facet.
A naive solution would be to compute similarity between the text embedding of facet-focused description $q_f$ and the image embedding of infographic $x_i$.
However, this method may not faithfully measure similarity along that specific facet, because different facets rely on different linguistic cues in the query and different visual evidence in the image.
We therefore seek facet-aware embeddings that model similarity separately for different facets.
To obtain them, we condition the text encoder with facet-specific tokens and project the image embedding through facet-specific heads, yielding facet-conditioned similarities.

\paragraph{Facet-aware text embeddings.} % using facet-specific tokens
We use a shared text encoder $E_T(\cdot)$ conditioned by facet-specific tokens to construct this.
Specifically, for each open-vocabulary facet $f$, we prepend a special token $\langle f\rangle$ to the facet rewrite $q_f$ to obtain the text embedding:
\begin{equation}
\label{eq:text-embed}
\mathbf{e}_{q,f} =
\begin{cases}
\mathrm{norm}\big(E_T(\langle f\rangle \oplus q_f)\big), & q_f\neq\emptyset,\\
\mathbf{0}, & q_f=\emptyset,
\end{cases}
\end{equation}
where $\mathrm{norm}(\cdot)$ denotes $\ell_2$ normalization, $\oplus$ denotes concatenation, and $\mathbf{0}$ is a zero vector of the corresponding dimensionality.

\paragraph{Facet-aware image embeddings.} % using facet-specific MLP heads
For each exemplar infographic $x_i$, we compute a base image embedding using an image encoder $E_I(\cdot)$:
$\mathbf{h}_{x_i}=\mathrm{norm}(E_I(x_i))$.
We then attach a lightweight MLP head per embedding facet to produce facet-specific image features:
\begin{equation}
\label{eq:image-embed}
\mathbf{e}_{x_i,f} = \mathrm{norm}\big(\mathrm{MLP}_f(\mathbf{h}_{x_i})\big).
\end{equation}
This multi-head design provides facet-conditioned projections of the same image representation, enabling different relevance notions for different facets.

\paragraph{Facet-conditioned similarities.}
For each embedding facet $f$, we compute cosine similarity between the text embedding $\mathbf{e}_{q,f}$ and the image embedding $\mathbf{e}_{x_i,f}$:
\begin{equation}
\label{eq:facet-sim}
s_f(q,x_i) = \mathbf{e}_{q,f}^\top \mathbf{e}_{x_i,f}.
\end{equation}
Please note that the facet-aware scoring function in Eq.~\ref{eq:facet-sim} is parameterized by 1) facet tokens $\{\langle f\rangle\}_{f\in\mathcal{F}_e}$ that condition the shared text encoder, and 2) facet-specific projection heads $\{\mathrm{MLP}_f\}_{f\in\mathcal{F}_e}$ that map a shared image embedding to facet spaces. 
These additional degrees of freedom are not explicitly supported by generic CLIP pretraining, especially for design-centric facets such as layout and style.
Therefore, we construct in-domain facet-level supervision and optimize the same facet-aware similarities, as described next.
% used at inference

% In practice, we precompute all image embeddings $\{\mathbf{e}_{x_i,f}\}_{i=1}^{N}$ for all the open-vocabulary facets and store them.
% At query time, we compute the text embeddings $\{\mathbf{e}_{q,f}\}$ and obtain all embedding-facet scores $\{s_f(q,x_i)\}_{i=1}^{N}$ efficiently via batched matrix multiplication between $\mathbf{e}_{q,f}$ and the stored embedding matrix for facet $f$.
% We compute $s_{\textsc{T}}(q,x_i)$ only when chart type is specified ($w_{\textsc{T}}>0$) and finally rank the corpus by Eq.~\ref{eq:final-score}, returning the top-$K$ exemplars.

\subsection{In-domain Alignment with Synthetic Facet Captions}
\label{sec:alignment}
To learn the facet-specific text token and MLP heads, we construct in-domain facet-level supervision from ChartGalaxy~\cite{li2026chartgalaxy} and perform contrastive alignment for each facet.

\paragraph{Alignment data construction.}
% We sampled 200k images from ChartGalaxy and removed near-duplicates in image-embedding space, yielding a training split of 52k images and a held-out test split of 10k images for alignment evaluation.
We removed near-duplicates in image-embedding space from the original ChartGalaxy dataset and randomly sampled 52,000 infographics for training.
For each training infographic $x_i$, we then generated facet-specific supervision for the four embedding facets by first producing a rich multimodal description with Qwen3-VL-8B and then distilling it into short facet-focused captions, following large-scale retrieval training practices such as MegaPairs \cite{zhou2025megapairs}.
For each open-vocabulary facet $f$, we created two paraphrased captions $\{c_f^{(1)}(x),c_f^{(2)}(x)\}$, yielding eight training texts per image, and trained on these distilled captions to provide cleaner facet-targeted supervision with less cross-facet leakage.

\paragraph{Multi-facet contrastive alignment.}
We optimize the same facet-aware similarity used at inference.
For each image $x$ in a minibatch $\mathcal{B}$, each facet $f$, and each caption variant $k\in\{1,2\}$,
we treat $(c_f^{(k)}(x),x)$ as a positive pair under Eq.~\ref{eq:facet-sim}$,$ and use other images in the batch as negatives:
\begin{equation}
\label{eq:facet-infonce}
\mathcal{L}=
\sum_{x\in\mathcal{B}}
\sum_{f\in\mathcal{F}_e}\sum_{k\in\{1,2\}}
-\log
\frac{\exp\big(\mathbf{e}_{c,f}^{(k)}(x)^\top\mathbf{e}_{x,f}/\tau\big)}
{\displaystyle\sum\limits_{x'\in\mathcal{B}} \exp\big(\mathbf{e}_{c,f}^{(k)}(x)^\top\mathbf{e}_{x',f}/\tau\big)} ,
\end{equation}
where $\mathbf{e}_{c,f}^{(k)}(x)=\mathrm{norm}\big(E_T(\langle f\rangle \oplus c_f^{(k)}(x))\big)$, $\tau$ is a temperature, and the image embedding $\mathbf{e}_{x,f}$ is computed via the same $\mathrm{MLP}_f$ heads in Eq.~\ref{eq:image-embed}.
This training encourages the model to learn facet-sensitive similarity functions that better reflect how users describe infographic exemplars.
% Chart type is not embedded or used in this alignment objective; it is handled as the discrete similarity term in Eq.~\ref{eq:chart-sim} at inference time.

% \paragraph{Updated parameters.}
% In the default setting, we update the facet-specific components (facet tokens and $\{\mathrm{MLP}_f\}$) while keeping the base encoders $E_T$ and $E_I$ fixed to preserve general-purpose alignment.
% Optionally, we further unfreeze $E_T$ and $E_I$ for a short second stage to better adapt the representation to infographic-specific language and design cues.

%%%%%%%%%%%%%%%%%%%%%%%%%%%%%%%%%%%%%%%%%%%%%%%%%%%%%%%%%%%%%%%%%%%%%%%%%%%%%%%%
\section{Conversational Retrieval-and-Adaptation System}
\label{sec:system}
\begin{figure*}[ht]
\centering
\includegraphics[width=\linewidth]{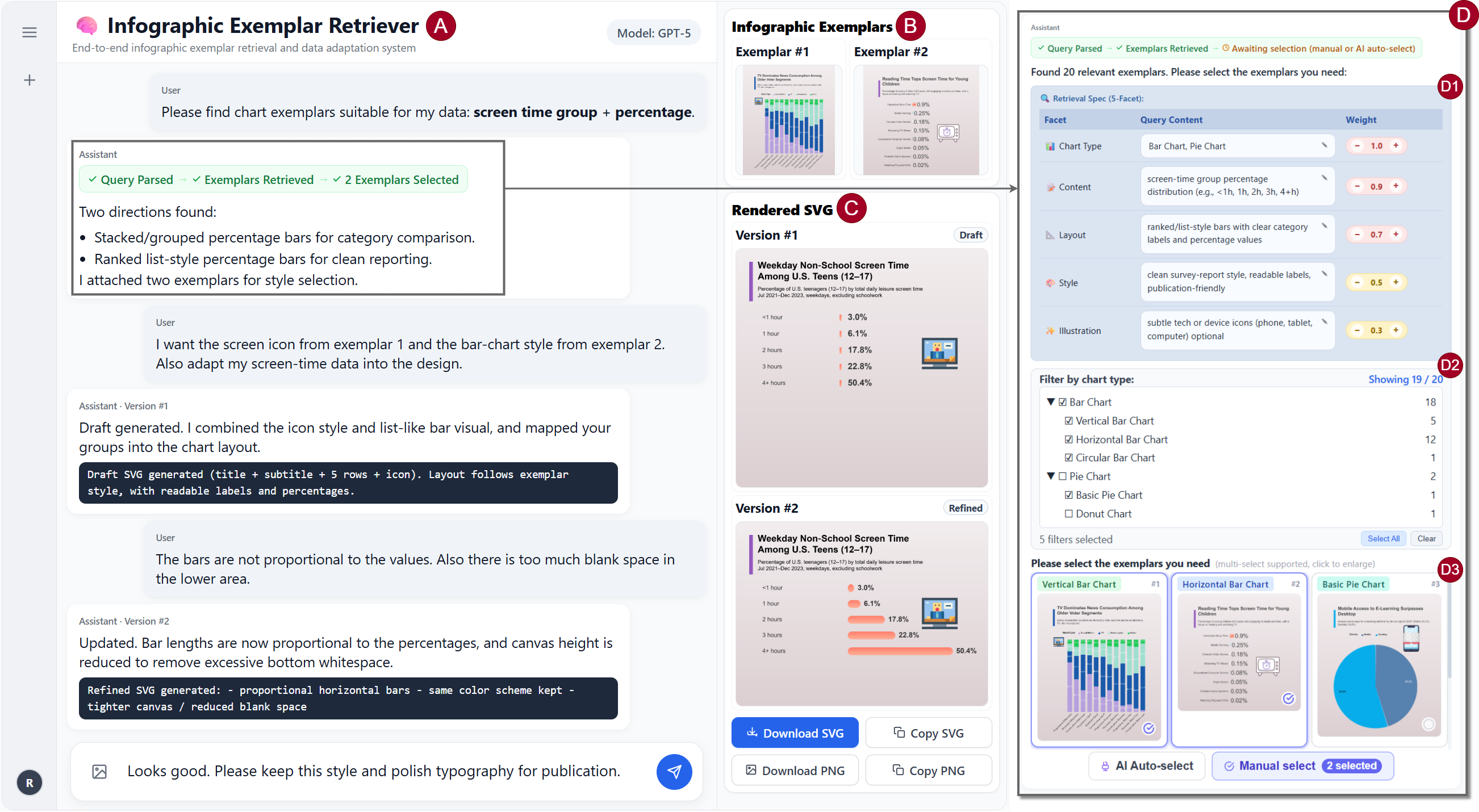}
\caption{\textbf{System UI for conversational exemplar retrieval and SVG-based adaptation.}
% It combines a chat-centered workflow with dedicated panels for exemplar retrieval and output inspection.
(A) chat panel for intent articulation, assistant feedback, and iterative refinement; (B) committed infographic exemplars that persist as references during adaptation; (C) rendered SVG outputs with version history; and (D) retrieval panel for exemplar discovery and commitment.
% The retrieval panel includes a facet-based retrieval specification with weights (D1), chart-type filters for narrowing candidates (D2), and a candidate gallery with multi-selection plus manual or AI auto-selection controls (D3).
}
\label{fig:system-ui}
\end{figure*}

Retrieval alone is often insufficient to support practical infographic authoring.
Even when a highly relevant exemplar is retrieved, users must still adapt its structure and visual style to their own data.
To bridge the gap between inspiration seeking and downstream reuse, we develop an exemplar-driven infographic authoring system, enabling users to 1) explore inspirations through retrieval, 2) commit to a small set of infographic exemplars, and 3) adapt those exemplars to their own data.

\subsection{Interface and Typical Workflow}

Figure~\ref{fig:system-ui} illustrates a concrete chat-centered workflow for exemplar retrieval and adaptation. In this example, the user first asks in the chat panel (A) for “chart exemplars suitable for my data.” The system parses this request into weighted retrieval facets shown in the retrieval panel (D1), supports chart-type filtering in (D2), and retrieves relevant exemplars in the candidate gallery (D3). After the user manually selects two exemplars, they remain pinned in the exemplar panel (B) as persistent references for downstream adaptation. The user then specifies a concrete combination strategy in the chat: keep the screen icon from exemplar~1, adopt the bar-chart style from exemplar~2, and adapt the design to the user’s screen-time data. The output panel (C) first shows a draft SVG (Version~\#1) that already integrates cues from both references, combining the illustration from one exemplar with the chart style of the other. After the user provides follow-up feedback to correct problems in the initial SVG output, the system generates a refined SVG (Version~\#2) with proportional horizontal bars and a tighter canvas while preserving the selected visual style. This example highlights how retrieval controls (D), pinned exemplars (B), conversational instructions (A), and iterative SVG refinement (C) work together in a single end-to-end authoring workflow.

% \subsection{System Architecture}
% \label{sec:system-arch}

% The backend integrates three components: (1) our infographic-exemplar retriever, (2) a single multimodal \emph{foundation model} that serves as the conversational backbone, and (3) a session store that persists conversation context and the committed exemplar set.
% The retriever implements the intent-aware scoring and ranking described in Section~\ref{sec:method}, surfacing candidate infographic exemplars for user exploration.
% The backbone model interprets user intent, orchestrates exemplar discovery and selection in the interface, and provides SVG-aware assistance grounded in the committed exemplars (including tool calls for node-level SVG inspection).

% To support the staged workflow, we invoke the backbone model with a small set of prompt templates aligned with the inspiration, commitment, and adaptation phases.

% \subsection{Commit: Infographic Exemplar Selection and Persistence}
% \label{sec:system-selection}

\subsection{Exemplar Retrieval}
\label{sec:system-retrieval}

During the inspiration phase, the assistant invokes the retriever described in Section~\ref{sec:method} to surface candidate infographic exemplars relevant to the user's request.
The retrieval panel (Figure~\ref{fig:system-ui}D) makes this process explicit by exposing an editable facet-based retrieval specification, together with hard chart-type filters and a browsable candidate gallery (D1--D3).
When users are not satisfied with the retrieval results, they can directly revise multi-facet queries and their importance weights in the panel, or continue refining the request through the chat, in which case the system updates the query text and weights and reruns retrieval.
After browsing the candidate gallery, users commit to a small set of infographic exemplars for downstream adaptation.
The system supports two selection modes: manual selection and AI auto-selection.
In AI auto-selection, a vision--language model re-ranks the top-10 retrieved candidates conditioned on the user's query and selects 1--3 exemplars, capturing fine-grained intent cues beyond similarity-based ranking alone while also promoting a more diverse set of references.
Once committed, the selected infographic exemplars persist across subsequent turns and are displayed alongside the conversation (B), grounding iterative follow-up requests in the same exemplar set.
% \subsection{Phase 2: SVG-Aware Assistance Grounded in Committed Infographic Exemplars}

\subsection{Infographic Adaptation}
\label{sec:system-svg}

After users commit to a small set of infographic exemplars, the system helps them adapt those exemplars to their own data.
Such adaptation typically involves directly fitting new tabular data into an existing design, reusing a specific visual style, or extracting particular visual elements (e.g., icons).
However, supporting this naively is difficult because infographic SVGs are often extremely long, containing embedded base64 assets and repetitive low-level structures that can easily overwhelm an MLLM's context window.
To address this, we design a progressive, tool-augmented adaptation pipeline driven by the user's data adaptation needs.

\paragraph{Default context: Compact structural summary.}
Whenever an exemplar is committed and the conversation enters the adaptation phase, the system does not feed the raw SVG to the model.
Instead, it automatically converts the SVG into a compact tree-structured summary that preserves hierarchical grouping (e.g., nested \texttt{<g>} elements) and coarse content signatures (e.g., dominant element types and counts).
Each node is assigned a stable \texttt{node\_id}.
This lightweight summary is provided to the assistant as the default context.
It enables the model to globally reason about \emph{where} a user's data should be mapped (e.g., marks, legends, or title regions) without incurring the cost of processing the full SVG source.

% For example, a compressed tree may summarize an infographic SVG as follows:

% \begin{svgbox}
% <svg> [node_1]
%   <defs> [node_2]
%     <linearGradient> [node_3]
%   <g class="chart"> [node_4]
%     <g class="bar-group"> [×3 similar] [node_5]
%       first instance:
%         <path class="bar"> [node_6]
%         <text class="value-label"> "32" [node_7]
%     <g class="legend"> [node_8]
%       <rect> [node_9]
%       <text> "Good" [node_10]
%   <g class="title"> [node_11]
%     <text> "Highest Good" [node_12]
%     <text> "Lowest Bad" [node_13]
%   <image class="decorative"> [node_14]
% \end{svgbox}

\paragraph{On-demand inspection: Sanitized code retrieval.}
When the user specifies a data adaptation request or a targeted styling change, the assistant determines which specific regions require modification based on the baseline structural summary.
It then invokes a system tool, \texttt{show\_full\_svg(node\_id)}, to retrieve the detailed, code-level representation for only those necessary nodes.
To further mitigate context bloat, the backend sanitizes this retrieved code on the fly: large embedded payloads like base64-encoded images are removed and replaced with lightweight placeholders.
This allows the assistant to zoom in and safely inspect layout-relevant SVG structure without processing massive binary strings.

\paragraph{Output generation and SVG reconstruction.}
With the necessary code snippets inspected, the assistant executes the adaptation---such as calculating new geometric attributes for data binding or updating text labels---by outputting the modified SVG code specifically for those queried \texttt{node\_id}s.
The system backend then automatically stitches these modified subtrees back into the overall SVG structure.
Crucially, during this reconstruction, the backend restores the original high-fidelity payloads, such as the base64 images, that were temporarily held out by placeholders.
The fully reconstructed SVG is then rendered in the output panel (Figure~\ref{fig:system-ui}C), seamlessly completing the adaptation loop and allowing the user to provide subsequent natural language commands for further refinement.
% \vicacomment{Still not clear how the adaptationresult can be generated.}
% \paragraph{Outputs.}
% Depending on user intent, the assistant produces (i) exemplar-grounded design guidance and (ii) concrete SVG code snippets or edits tied to the inspected nodes.
% The interface renders SVG outputs for quick verification and supports iterative follow-up requests (e.g., ``use percentage labels'', ``align tick labels'', ``highlight the maximum'') grounded in the same committed exemplars.

%%%%%%%%%%%%%%%%%%%%%%%%%%%%%%%%%%%%%%%%%%%%%%%%%%%%%%%%%%%%%%%%%%%%%%%%%%%%%%%%
\begin{table*}[t]
  \centering
  \scriptsize
  \caption{\textbf{Single-round query-to-image retrieval.} We report Recall@1/5 (\%) and MRR@10 (fraction) on synthetic general queries, synthetic multi-facet queries, and human-written paired short/long queries.
  $\uparrow$ higher is better.}
  \label{tab:single-round}
  \setlength{\tabcolsep}{3.2pt}
  \begin{tabular}{p{0.2\textwidth}ccc|ccc|ccc|ccc}
  \toprule
  \textbf{Method} &
  \multicolumn{3}{c|}{\textbf{Synthetic (General)}} &
  \multicolumn{3}{c|}{\textbf{Synthetic (Multi-facet)}} &
  \multicolumn{3}{c|}{\textbf{Human (Short)}} &
  \multicolumn{3}{c}{\textbf{Human (Long)}} \\
  \cmidrule(lr){2-4}\cmidrule(lr){5-7}\cmidrule(lr){8-10}\cmidrule(lr){11-13}
  & \textbf{R@1}$\uparrow$ & \textbf{R@5}$\uparrow$ & \textbf{MRR@10}$\uparrow$
  & \textbf{R@1}$\uparrow$ & \textbf{R@5}$\uparrow$ & \textbf{MRR@10}$\uparrow$
  & \textbf{R@1}$\uparrow$ & \textbf{R@5}$\uparrow$ & \textbf{MRR@10}$\uparrow$
  & \textbf{R@1}$\uparrow$ & \textbf{R@5}$\uparrow$ & \textbf{MRR@10}$\uparrow$ \\
  \midrule
  CLIP~\cite{radford2021clip} & 62.47 & 84.74 & 0.7203 & 46.42 & 70.83 & 0.5696 & 35.00 & 58.00 & 0.4375 & 41.00 & 67.33 & 0.5150 \\
  SigLIP2~\cite{tschannen2025siglip} & 19.38 & 38.05 & 0.2739 & 67.59 & 86.45 & 0.7566 & 33.00 & 50.00 & 0.4042 & 20.67 & 33.33 & 0.2632 \\
  MegaPairs~\cite{zhou2025megapairs} & 67.42 & 88.18 & 0.7630 & 50.08 & 74.87 & 0.6075 & 39.33 & 65.67 & 0.5036 & 50.33 & 73.67 & 0.5913 \\
  \midrule
  \textsc{Ours w/o In-dom. Align.} & 70.22 & 87.43 & 0.7760 & 51.38 & 75.76 & 0.6168 & 30.00 & 47.00 & 0.3732 & 35.00 & 57.33 & 0.4424 \\
  \textsc{Ours w/o Facets} & 92.79 & 98.88 & 0.9545 & 89.67 & 97.60 & 0.9332 & 53.67 & 77.33 & 0.6345 & 65.00 & 82.33 & 0.7237 \\
  \textsc{Ours w/o Facet Weights} & \textbf{95.92} & \textbf{99.72} & \textbf{0.9760} & 90.85 & 98.40 & 0.9422 & 54.33 & 81.67 & 0.6552 & 69.33 & \textbf{86.67} & 0.7689 \\
  \textsc{Ours} & 95.72 & 99.68 & 0.9746 & \textbf{91.29} & \textbf{98.48} & \textbf{0.9447} & \textbf{54.67} & \textbf{82.00} & \textbf{0.6591} & \textbf{70.33} & \textbf{86.67} & \textbf{0.7744} \\
  \bottomrule
  \end{tabular}
\end{table*}
\section{Experiments}
\label{sec:evaluation}

We evaluate both our retriever and the end-to-end authoring system in three complementary settings: 1) single-round retrieval, 2) multi-round retrieval, and 3) retrieval-based authoring.

\subsection{Single-Round Retrieval}
\label{sec:eval-single-round}

\subsubsection{Automatic Retrieval Evaluation}

We randomly select 10,000 infographics from the ChartGalaxy to form a fixed corpus $\mathcal{X}$.
Given each pair of a text query $q$ and a unique ground-truth target infographic $x^{*}\in\mathcal{X}$, a retriever ranks all infographics in $\mathcal{X}$ and returns a top-$K$ list.
We report Recall@K and MRR@10 to evaluate the retrieval performance under a unique-target protocol.

\paragraph{The construction of query-target pairs.}
We construct three types of query-target pairs to cover different levels of query specificity.\looseness=-1
\begin{itemize}[nosep]
  \item \textbf{Synthetic general queries.}
  For each target infographic $x_i \in \mathcal{X}$, we generate a rich natural-language description by prompting Qwen3-VL to summarize its visual content.
  This description is directly used as the retrieval query $q_i$.
  This serves as a natural synthetic baseline for measuring whether the retriever can align full-image descriptions with the corresponding visual instances.
  
  \item \textbf{Synthetic multi-facet queries.}
  Building on the rich description above, we further generate its facet-specific descriptions using the same pipeline described in \Cref{sec:method}. % $\{q_f^{(i)}\}_{f\in\{\textsc{C},\textsc{L},\textsc{I},\textsc{S}\}}$
  We then construct a composite query $q_i$ by concatenating these facet-specific descriptions.
  This is designed to test the retriever's ability to leverage more structured and fine-grained design intent specifications.
%together with the ground-truth chart type.
  \item \textbf{Human-written queries.}
  To evaluate retrieval under more realistic user-facing conditions and avoid potential biases in the synthetic query generation, we randomly sample 300 infographics from $\mathcal{X}$ and collect human-written queries.
  Annotators were instructed to write both a short query and a long query for each infographic.
  The short query captures only the most salient cues required to identify the intended infographic, while the long query provides a more complete specification by describing additional aspects, such as layout and style, in free-form natural language.

  The detailed prompt and instructions to the annotators are provided in Supplementary Section~5.2.
\end{itemize}

% \paragraph{Human query collection protocol.}
% Annotators were instructed to describe the intended exemplar in their own words and to avoid copying visible on-image text verbatim (e.g., titles or labels).
% When textual content was essential to express intent, annotators were encouraged to paraphrase rather than quote.
% Queries were written to be self-contained and to focus on retrievable cues such as content, chart type, layout, and illustration/style attributes.
% Please see supplementary for full instructions and examples.

\paragraph{Baseline methods and ablations.}
We compare our method (\textsc{Ours}) with the following baselines:

\begin{itemize}[noitemsep]
  \item \textbf{CLIP}~\cite{radford2021clip}: a standard and widely used CLIP ViT-B/32 dual-encoder baseline that well captures text--image similarity. % (implemented using the \texttt{clip-vit-base-patch32} checkpoint)
  \item \textbf{SigLIP2}~\cite{tschannen2025siglip}: a more recent and stronger dual-encoder baseline. % (implemented using the \texttt{siglip2-base-patch16-224} checkpoint).
  \item \textbf{MegaPairs}~\cite{zhou2025megapairs}: a retrieval-oriented model trained with large-scale synthetic supervision. % dual-encoder
\end{itemize}

We also consider the following ablations to examine the contribution of individual components
\begin{itemize}[noitemsep]

  \item \textbf{\textsc{Ours w/o Facets}}: removes facet decomposition and query-dependent fusion. Specifically, it only rewrites each query and then retrieves infographics using the same model as our method.
  \item \textbf{\textsc{Ours w/o Facet Weights}}: uses uniform weights over present embedding facets.
  \item \textbf{\textsc{Ours w/o In-domain Alignment}}: removes the in-domain alignment step in our training pipeline.
  % \item \textbf{\textsc{Ours w/o Chart Type}}: sets $w_{\textsc{T}}{=}0$ for all queries, \ie, removes the discrete chart type.
\end{itemize}

% \paragraph{Baseline evaluation setting.}
% All external baselines (CLIP, SigLIP2, MegaPairs) are evaluated \emph{as-is}, without any additional in-domain fine-tuning on our infographic corpus.
% We follow each model's recommended preprocessing to compute a single text embedding for the input query and a single image embedding for each corpus image, and we rank images by the model's default text--image similarity.
% For B1, the baseline input is the composite query formed by concatenating the four facet descriptions into a single text string.
% For B2, the baseline input is the original short/long user-written query text.
% This ensures that baselines operate on exactly the same textual inputs provided by each benchmark, without access to our facet decomposition or query-dependent fusion.

\paragraph{Metrics.}
Let $\mathrm{rank}(q)$ be the rank of the target infographic for query $q$, and $\mathcal{Q}$ be the set of all queries. % 1-indexed 
We report Recall@K and MRR@10.
% Unless noted otherwise, we report \textbf{Recall@K in \%} and \textbf{MRR@10 as a fraction in $[0,1]$}.

\textbf{Recall@K (R@K)} is the fraction of queries whose target appears in the top-$K$ ranked list:
\begin{equation}
\mathrm{R@}K = \frac{1}{|\mathcal{Q}|}\sum_{q\in\mathcal{Q}} \mathbf{1}\big[\mathrm{rank}(q)\le K\big],
\quad K\in\{1,5\}.
\end{equation}

% \textbf{MRR@10.}
% We report mean reciprocal rank truncated at 10:
\textbf{MRR@10} is the mean reciprocal rank of the ground-truth target, truncated at the top 10 results:
\begin{equation}
\mathrm{MRR@10} = \frac{1}{|\mathcal{Q}|}\sum_{q\in\mathcal{Q}}
\begin{cases}
\frac{1}{\mathrm{rank}(q)}, & \mathrm{rank}(q)\le 10,\\
0, & \text{otherwise}.
\end{cases}
\end{equation}

\paragraph{Results.}
\Cref{tab:single-round} summarizes single-round retrieval results. %  on different types of queries
% We treat the human-written benchmark as the primary indicator of intent alignment under realistic authoring-oriented queries.
Overall, \textsc{Ours} consistently improves retrieval compared to strong off-the-shelf retrievers, with especially large gains on human-written queries.
The in-domain alignment step is necessary for improving retrieval performance, and explicitly modeling multiple facets can further improve retrieval performance, especially for long human-written queries.
% Comparing \textsc{Ours} to \textsc{Ours w/o In-dom Align} shows the necessity of the in-domain alignment step for improving retrieval performance, and comparing \textsc{Ours} to \textsc{Ours w/o Facets} highlights the benefit of explicitly modeling multiple facets rather than collapsing the entire request into a single rewritten query.

A notable pattern among the external baselines is SigLIP2~\cite{tschannen2025siglip}, which remains competitive on synthetic multi-facet queries, but drops significantly on synthetic general queries and long human-written queries.
This suggests that a global text embedding is less effective when multiple facets are interleaved. %  in one long description
Our method overcomes this by explicitly modeling multiple facets and then performing retrieval.
% By decomposing and scoring these facets separately, \textsc{Ours} is better aligned with long design-intent queries.
% Comparing \textsc{Ours} to \textsc{Ours w/o Facets} further highlights the benefit of explicitly modeling multiple facets (content/layout/illustration/style) rather than collapsing the entire request into a single rewritten query.

% A notable pattern among the external baselines is SigLIP2~\cite{tschannen2025siglip}: it remains competitive on structured synthetic multi-facet queries, but drops on synthetic rich captions and human long queries. This pattern suggests that a single global text embedding is less effective when content, layout, style, and illustration cues are interleaved in one long description. By decomposing and scoring these facets separately, \textsc{Ours} is better aligned with long design-intent queries.

\subsubsection{Human Judgment of Retrieved Results}
\label{sec:eval-human-judgment}

% Automatic metrics under a unique-target protocol may not fully reflect the usefulness of retrieved exemplars for authoring.
% We therefore conduct a human scoring study comparing retrieval outputs from MegaPairs and \textsc{Ours} on the human-written queries (B2).
Automatic metrics under a unique-target protocol may not fully reflect retrieval usefulness for infographic authoring, because a query can correspond to multiple highly relevant exemplars.
To better capture practical usefulness, we complemented automatic evaluation with a human judgment study on the retrieved lists. %, comparing MegaPairs and \textsc{Ours} on the human-written queries.

\paragraph{Evaluation setup.}
For each human-written query, we retrieve the top-$5$ results returned by MegaPairs and \textsc{Ours}, respectively.
These two ranked lists are presented side by side in randomized order.
Each query--method pair is scored independently by two raters, and we use the average of the two scores in all analyses.
Raters score each list on a 1--5 Likert scale based on overall intent match and usefulness as an exemplar, where 1 indicates \emph{irrelevant} and 5 indicates \emph{near-perfect match}.
To anchor the upper end of the scale, raters are instructed to assign a score of 5 when the target or intended exemplar already appears at rank~1, since the retrieval has then fully satisfied the user's need.
% Three expert raters score each list on a 1--5 Likert scale based on overall intent match and usefulness as an exemplar:
% (1) irrelevant, (2) weak match, (3) partially useful, (4) strong match, and (5) near-perfect match.

\paragraph{Results.}
We report the mean of the two raters' scores for each query--method pair, and summarize win/tie/loss using paired comparisons of these averaged scores.
Inter-rater agreement was high, with a quadratic weighted Cohen's $\kappa$ of 0.98 and 90.0\% exact agreement.
% This level of agreement is consistent with our explicitly anchored rubric: when the intended exemplar is ranked at top~1, raters assign a score of 5, which reduces ambiguity in clearly successful cases.
As shown in \Cref{tab:human-score}, \textsc{Ours} substantially outperforms MegaPairs in perceived exemplar usefulness, increasing the mean score from 3.09 to 4.20 for short queries, from 3.26 to 4.51 for long queries, and from 3.17 to 4.35 overall.
The larger improvement on long queries suggests that \textsc{Ours} better exploits richer intent descriptions beyond topical similarity.
Across 600 paired comparisons, \textsc{Ours} wins in 57.3\% of cases, ties in 37.3\%, and loses in only 5.3\%.
% We report mean score computed per query by comparing the two systems' scores for short queries, long queries, and overall.
% It 

% as well as per-query win/tie/loss rates 

% Human judgments show a large gap in perceived exemplar usefulness in favor of \textsc{Ours} (\Cref{tab:human-score}).
% Beyond the mean-score lift, two patterns are informative.
% First, the improvement is stronger on \emph{long} queries (4.48 vs.\ 3.21), suggesting that \textsc{Ours} better capitalizes on richer intent descriptions that include layout/structure and style/illustration cues, rather than relying primarily on topical similarity.
% Second, the relatively high tie rate (40\%) indicates that for a substantial subset of queries, both systems can surface at least one partially-to-strongly useful exemplar in the top-$5$;
% however, \textsc{Ours} more frequently upgrades these cases into \emph{strong} or \emph{near-perfect} matches (win rate 54.9\% with only 5.1\% losses), which aligns with the goal of reducing ``near-miss'' results that match content but miss key presentation attributes important for authoring.

\begin{table}[t]
\centering
\scriptsize
\caption{\textbf{Human scoring of top-5 retrieved results.} $\uparrow$ higher is better.
% \emph{Win/Tie/Loss} is computed by comparing these paired averaged scores over \textbf{600} human-written queries (300 targets $\times$ \{short,long\});
% each query compares two top-$5$ result lists.
}
\label{tab:human-score}
\setlength{\tabcolsep}{4pt}
\begin{tabular}{lccc|ccc}
\toprule
& \multicolumn{3}{c|}{\textbf{Mean score} (1--5)$\uparrow$} & \multicolumn{3}{c}{\textbf{Preference vs.\ MegaPairs} (\%)} \\
\cmidrule(lr){2-4}\cmidrule(lr){5-7}
\textbf{Method} & \textbf{Short} & \textbf{Long} & \textbf{Overall}
& \textbf{Win} & \textbf{Tie} & \textbf{Loss} \\
\midrule
MegaPairs & 3.09 & 3.26 & 3.17 & -- & -- & -- \\
\textsc{Ours} & \textbf{4.20} & \textbf{4.51} & \textbf{4.35} & \textbf{57.3} & 37.3 & 5.3 \\
\bottomrule
\end{tabular}
\end{table}

%%%%%%%%%%%%%%%%%%%%%%%%%%%%%%%%%%%%%%%%%%%%%%%%%%%%%%%%%%%%%%%%%%%%%%%%%%%%%%%%
\subsection{Multi-Round Retrieval}
\label{sec:eval-img-guided-caption}
Beyond the single-round evaluation, we conducted a user study to assess retrieval performance in an interactive multi-round setting.
% The goal was to test whether our method helps users converge more efficiently to useful exemplars while iteratively refining caption-style queries with the target image kept in view.
% This study evaluates whether our method helps users converge more efficiently to useful exemplars during iterative search.
% We focus on a multi-round setting in which participants repeatedly refine caption-style queries while seeing a visual target, allowing us to measure both retrieval effectiveness and interaction efficiency.

\subsubsection{Study Design}
\label{sec:img-guided-design}

\paragraph{Participants.}
We recruited 12 participants (P1-P12, 9 male, 3 female), age 20--25 ($M{=}22.67$, $SD{=}1.37$).
Four identified as design novices (33.3\%), six reported some prior experience (50.0\%), and two reported proficiency (16.7\%).
Participants reported moderate-to-high experience in graphic design or layout tools ($M{=}4.08/5$, $SD{=}1.24$).

\paragraph{Task and stimuli.}
The task is to retrieve examples that are semantically and visually similar to the target infographic.
We selected four representative infographics from the corpus $\mathcal{X}$ to cover diverse visual styles, including variation in chart type, layout structure, illustration density, and overall appearance.
In each task, participants were shown one target infographic and used a web interface consisting of a target image panel, a text input box, a top-$10$ result grid, and a save panel for collecting candidate results.
Participants can iteratively write and revise the query, save promising results throughout the search, rate saved results, and stop whenever they feel ready to choose a best result.
% After each submission, the system refreshed the top-$10$ results; we refer to each submission-and-refresh cycle as one \emph{round}.
% Participants can 

\paragraph{Conditions.}
We compared \textsc{Ours} with a \textsc{Baseline} that uses the same LLM as our system (\textbf{GPT-5}) to rewrite the participant's current input into a single holistic caption-style query, followed by retrieval with MegaPairs~\cite{zhou2025megapairs}.
In contrast to \textsc{Ours}, the baseline does not construct a five-facet retrieval specification and performs retrieval from only one rewritten query.
The study used a counterbalanced within-subject design.
Specifically, each participant completed four tasks, with two targets assigned to \textsc{Ours} and two to \textsc{Baseline}.

% One round consists of a single query submission and the subsequent refresh of the retrieved results. Each participant completes four tasks (one per target), with two targets assigned to each condition; assignments are counterbalanced across participants so that every target is evaluated under both systems.

% \paragraph{Conditions.}
% We compare \textsc{Ours} against a \textsc{Baseline} that uses the same LLM as our system (\textbf{GPT-5}) to rewrite the participant's current input into a single holistic caption-style query, and then retrieves with MegaPairs~\cite{zhou2025megapairs}.
% Unlike \textsc{Ours}, this baseline does not construct a five-facet retrieval specification and performs retrieval from one rewritten query only.
% Each participant completes four tasks (one per target).
% Within each participant, two targets are assigned to \textsc{Ours} and two to \textsc{Baseline}.
% Across participants, assignments are counterbalanced so that each target is evaluated under both systems.

% The study used a within-subjects design: each participant completed tasks under both \textsc{Ours} and \textsc{Baseline}, with target--condition assignments counterbalanced across participants.

\paragraph{Study procedure.}
Each session began with a brief introduction to the task and interface, followed by one practice task to familiarize participants with the multi-round retrieval workflow.
Participants then completed four formal tasks.
During each task, participants were required to save at least two retrieved infographics that they considered good matches or useful alternatives.
For every saved infographic, they provided a satisfaction rating on a 1--7 Likert scale to indicate how well it matches the target infographic.
When ready to stop, they selected one saved infographic as the \textbf{best} result for the task.
We marked a task as \textbf{Found} if the participant saved the exact target infographic.
The full session took approximately 30 minutes.

\subsubsection{Metrics}
\label{sec:img-guided-metrics}

We use four metrics to evaluate the performance of multi-round retrieval.
\textbf{Rounds-to-stop} measures how many interaction rounds a participant completed before deciding to stop.
\textbf{FoundRate} is the proportion of tasks in which the participant saved the exact target infographic.
\textbf{dCRR@10} is a discounted cumulative reciprocal-rank measure adapted from CRR~\cite{chaney2015social}.
Specifically, it first computes the per-round reciprocal rank truncated at 10 and then discounts later rounds by a factor of $\gamma$:
\[
  \mathrm{dCRR@10}
  = \sum_{r=1}^{R} \gamma^{r-1}\,\mathrm{RR@10}_{r},
\]
where $R$ is the rounds-to-stop, $\gamma=0.9$ is the discount factor,
and $\mathrm{RR@10}_{r}$ is the reciprocal rank truncated at 10 in round $r$.
Thus, dCRR@10 rewards finding the target earlier and ranking it higher.
\textbf{BestSatisfaction} is the 1--7 satisfaction rating assigned to the participant's final selected image.

\subsubsection{Statistical Analysis}
\label{sec:img-guided-analysis}

Each participant completed four tasks, two under each condition, yielding 48 task observations in total.
For inferential testing, we first aggregated task-level outcomes within each participant and condition, producing one paired value per participant for each metric: mean Rounds-to-stop, mean dCRR@10, mean BestSatisfaction, and the participant-level proportion for FoundRate.
We then compared conditions using paired two-sided $t$-tests across participants.
Reported $p$ values are Holm--Bonferroni corrected across the four reported metrics.
Descriptive statistics in \Cref{tab:img-guided-results} are shown at the task level.

\begin{table}[t]
\centering
\small
\caption{
\textbf{Multi-round retrieval.}
We report task-level mean$\pm$SD for continuous metrics and percentage (count) for binary metrics.
$\downarrow$ lower is better; $\uparrow$ higher is better.
% For inference,
% $p$ values are from participant-level paired two-sided $t$-tests and are Holm--Bonferroni corrected. % across the four reported metrics.
}
\label{tab:img-guided-results}
\setlength{\tabcolsep}{6pt}
\renewcommand{\arraystretch}{1.03}
\begin{tabular}{@{}p{1.45cm}p{1.90cm}ccc@{}}
\toprule
\textbf{Category} & \textbf{Metric} & \textbf{\textsc{Baseline}} & \textbf{\textsc{Ours}} & \textbf{$p$} \\
\midrule
Efficiency
& Rounds-to-stop $\downarrow$
& 3.29$\pm$2.27
& \textbf{1.42$\pm$0.50}
& 0.0029 \\
\midrule

\multirow[t]{2}{1.45cm}{\raggedright Retrieval}
& FoundRate $\uparrow$
& 45.8\% (11/24)
& \textbf{91.7\% (22/24)}
& 0.0018 \\
& dCRR@10 $\uparrow$
& 0.21$\pm$0.34
& \textbf{0.73$\pm$0.38}
& $5.7\times10^{-5}$ \\
\midrule

\parbox[t]{1.45cm}{\raggedright Utility (1--7)}
& BestSatisfaction $\uparrow$
& 5.50$\pm$1.62
& \textbf{6.88$\pm$0.45}
& 0.0029 \\
\bottomrule
\end{tabular}
\end{table}

\subsubsection{Results}
\label{sec:img-guided-results}

\Cref{tab:img-guided-results} summarizes the primary results.
Overall, \textsc{Ours} improves exact-match retrieval and user utility while reducing interaction rounds.
Compared to the single-query rewrite baseline, \textsc{Ours} reduces Rounds-to-stop by 1.87 rounds (3.29 vs.\ 1.42; $p{=}0.0029$).
Exact-match success nearly doubles, increasing from 45.8\% (11/24) to 91.7\% (22/24), a gain of 45.9 percentage points ($p{=}0.0018$).
We also observe substantially stronger cumulative exposure across rounds, with dCRR@10 increasing from 0.21 to 0.73 (+0.52; $p{=}5.7\times10^{-5}$), and higher satisfaction for the final selected result, with BestSatisfaction increasing from 5.50 to 6.88 (+1.38; $p{=}0.0029$).

%%%%%%%%%%%%%%%%%%%%%%%%%%%%%%%%%%%%%%%%%%%%%%%%%%%%%%%%%%%%%%%%%%%%%%%%%%%%%%%%

%%%%%%%%%%%%%%%%%%%%%%%%%%%%%%%%%%%%%%%%%%%%%%%%%%%%%%%%%%%%%%%%%%%%%%%%%%%%%%%%
%%%%%%%%%%%%%%%%%%%%%%%%%%%%%%%%%%%%%%%%%%%%%%%%%%%%%%%%%%%%%%%%%%%%%%%%%%%%%%%%
\subsection{Retrieval-based Authoring}
\label{sec:eval-end-to-end}

Beyond evaluating retrieval quality alone, we further assess the full authoring workflow enabled by our system.
% Specifically, we assess whether integrating exemplar retrieval, exemplar selection, and SVG adaptation in a unified interface better supports infographic authoring than a separated toolchain.
% \Cref{fig:system-ui} shows a representative System~A session of this process.
% In that case, the user first requests exemplars for a ``screen time group + percentage'' dataset, reviews two retrieved exemplars, and then asks the system to combine the screen icon from exemplar~1 with the bar-chart style from exemplar~2.
% The system generates an SVG draft, receives follow-up feedback about bar proportionality and excessive blank space, and then returns a refined version.

% The retrieval evaluations above ask whether the system can surface relevant exemplars.
% We next evaluate the \emph{full authoring workflow}: whether integrating exemplar retrieval, exemplar selection, and SVG adaptation in a single interface helps users complete infographic-authoring tasks with lower overhead than a separated toolchain.
% \Cref{fig:system-ui} shows a representative System~A session of the same workflow.
% In that case, the user first requests exemplars for a ``screen time group + percentage'' dataset, reviews two retrieved exemplars, and then asks the system to combine the screen icon from exemplar~1 with the bar-chart style from exemplar~2.
% The system generates an SVG draft, receives follow-up feedback about bar proportionality and excessive blank space, and then returns a refined version.
% This search--select--adapt--refine loop is the interaction pattern evaluated in this study.
\subsubsection{Study Design}

\paragraph{Participants.}
The same 12 participants (P1-P12) from the multi-round search study (\Cref{sec:eval-img-guided-caption}) completed this end-to-end evaluation.

\paragraph{Task and stimuli.}
Each participant completed two infographic-authoring tasks using two tabular datasets from different topical domains. 
Dataset A captured commuter mode share in Champaign County, Illinois, as percentages across seven travel modes. Dataset B captured weekday daily screen time among U.S. teenagers ages 12--17, grouped into five duration categories excluding schoolwork.
% \vicacomment{Briefly describe the datasets.}
% Original: In each task, participants were given a tabular dataset and asked to retrieve suitable infographic exemplars, adapt one or more selected exemplars to the data, and iteratively refine the generated SVG until satisfied or unable to make further progress.
For each task, participants were asked to retrieve infographic exemplars, adapt one or more selected exemplars to the data, and iteratively refine the generated SVG until they were satisfied or could no longer make further progress.

%\textbf{counterbalanced} dataset assignment across systems (half of participants used dataset~A with System~A and dataset~B with System~B; the other half used the reverse assignment).
% We did not impose a hard time limit.

% : one with our integrated system (\textbf{System A}) and one with a toolchain baseline (\textbf{System B})
\paragraph{Conditions.}
% Original: \textbf{System A} is our one-stop interface that integrates exemplar retrieval, exemplar selection, and iterative SVG adaptation with a live preview and persistent exemplars (\Cref{sec:system}).
\textbf{\textsc{Ours}} is our one-stop interface that integrates exemplar retrieval, exemplar selection, and iterative SVG adaptation with a live preview and persistent exemplars (\Cref{sec:system}).
% Original: \textbf{System B} is a toolchain baseline that uses the same LLM as System~A (\textbf{GPT-5}) but exposes the workflow as separate tools:
\textbf{Baseline} is a toolchain baseline that uses the same LLM as \textsc{Ours} (\textbf{GPT-5}) but exposes the workflow as separate tools:
(i) a MegaPairs retriever~\cite{zhou2025megapairs} for finding exemplars,
(ii) a plain chat interface for asking GPT-5 to adapt and edit SVG code, and
(iii) an external SVG viewer for previewing intermediate results.
% Original: This condition serves as a realistic toolchain baseline: participants still had access to a strong retriever and the same backbone model, but they had to manually transfer selected exemplars and SVG code between tools, and they did not have a unified interface for exemplar management and iterative preview.
This condition serves as a realistic toolchain baseline, where participants could use MegaPairs retriever and backbone model, but had to manually transfer exemplars and SVG code across tools. % without a unified interface for exemplar management and iterative preview
We counterbalanced the system order for each participant.

\paragraph{Study procedure.}
% System order was \textbf{counterbalanced} (half used A$\rightarrow$B; half used B$\rightarrow$A).
At the beginning of the session, participants were introduced to the experimental setup and given instructions on how to interact with the two systems.
During each session, participants completed two assigned tasks, using one system for each task.
The experimenter observed the interaction process and maintained a structured record sheet to document process-level observations, including the number of dialogue turns, whether the final output satisfied the task requirements, and any notable breakdowns or recovery behaviors that occurred during task completion.
After completing the tasks, participants filled out a post-study questionnaire and provided open-ended feedback.
In addition, we conducted informal follow-up questions with four participants to clarify their experiences.

% \paragraph{Measures.}
% % \textbf{Workload (NASA-TLX).}
% We measured perceived workload using the NASA Task Load Index (NASA-TLX)~\cite{hart1988development}.
% As our authoring tasks involved negligible physical demand and self-rated performance was assessed separately, we included four NASA-TLX subscales: \emph{mental demand}, \emph{temporal demand}, \emph{effort}, and \emph{frustration}.
% Each subscale was rated on a a 21-point scale from 0 to 20. (0=very low; 20=extremely high).
% Given the small sample size, we treat the aggregate \textbf{raw NASA-TLX workload score} (\textbf{RTLX}), computed as the average of the four subscales, as the primary quantitative endpoint and report item-level metrics descriptively.
% Because each participant used both systems, we evaluated differences in workload outcomes using paired two-sided $t$-tests.

% % \textbf{Satisfaction.}
% Participants also rated three satisfaction dimensions on a 21-point scale from 0 to 20: self performance, tool satisfaction, and output satisfaction.
% For these items, 0 indicated \emph{not satisfied at all} and 20 indicated \emph{extremely satisfied}.
% We report item-level scores and an aggregate \textbf{Satisfaction} score (average of the three items), which are also analyzed using paired two-sided $t$-tests.

\subsubsection{Measures}
We measured perceived workload, satisfaction, and output quality.

\textbf{Perceived workload} was assessed using the NASA Task Load Index (NASA-TLX)~\cite{hart1988development}.
Because the authoring tasks involved minimal physical demand, we included four NASA-TLX subscales: \emph{mental demand}, \emph{temporal demand}, \emph{effort}, and \emph{frustration}.
Each subscale was rated on a 21-point scale from 0 (\emph{very low}) to 20 (\emph{extremely high}).
We computed the mean of them as the primary workload outcome (RTLX).
% Original: We used the aggregate raw NASA-TLX score (RTLX), calculated as the mean of the four subscales, as the primary workload outcome, while reporting subscale-level results descriptively.
% We used the aggregate raw NASA-TLX score (RTLX), defined as the mean of the four subscales, as the primary workload outcome.

\textbf{Satisfaction} was measured on the same 21-point scale across three dimensions: self-rated performance, tool satisfaction, and output satisfaction, where 0 indicated \emph{not satisfied at all} and 20 indicated \emph{extremely satisfied}.
We also computed the mean of them (Satisfaction). %as the primary satisfaction outcome 
% We report both item-level scores and an aggregate \textbf{Satisfaction} score, computed as the mean of the three items, as secondary outcomes, while reporting subscale-level results descriptively.
% Because each participant used both systems, differences in workload and satisfaction were analyzed using paired two-sided $t$-tests.

\textbf{Output quality} was assessed through a blind peer-review protocol among the participants.
Each participant evaluated 4 pairs of final infographics produced by other participants.
% After all authoring sessions were completed, each of the 12 participants was asked to evaluate 4 pairs of final infographics produced by other participants.
% Participants were explicitly excluded from evaluating their own outputs.
For each pair, the two infographics were generated using the exact same tabular dataset but with different systems by different authors.
Participants were asked to judge which output was better overall or if they were of equal quality. % (Left better, Right better, or Tie)
This procedure resulted in $12\times 4 = 48$ pairwise judgments. % ($12 \text{ participants} \times 4 \text{ pairs}$)
% We summarize these judgments to compare the overall preference distribution between different systems.

% The system identities and author names were strictly hidden, and the left/right presentation order was randomized.
\begin{table}[t]
\centering
\small
% Original: \caption{\textbf{End-to-end authoring study.}
% Original: We report mean$\pm$SD for continuous metrics.
% Original: Questionnaire ratings use a 0--20 scale.
% Original: Workload items follow NASA-TLX subscales; \textit{overall workload} is RTLX, computed as the unweighted mean of mental demand, temporal demand, effort, and frustration.
% Original: Lower is better for $\downarrow$ metrics; higher is better for $\uparrow$ metrics.
% Original: $p$ values are from paired two-sided $t$-tests.}
\caption{\textbf{Retrieval-based authoring.}
We report mean$\pm$SD for continuous metrics and percentage for preference rate.
% Questionnaire ratings use a 0--20 scale.
% Original: Lower is better for $\downarrow$ metrics; higher is better for $\uparrow$ metrics.
$\downarrow$ lower is better; $\uparrow$ higher is better.
$p$ values are from paired two-sided $t$-tests.}
\label{tab:e2e-authoring}
\setlength{\tabcolsep}{5pt}
\renewcommand{\arraystretch}{1.03}
\begin{tabular}{@{}p{1.45cm}p{2.35cm}ccc@{}}
\toprule
\textbf{Category} & \textbf{Metric} & \textsc{Baseline} & \textsc{Ours} & \textbf{$p$} \\
\midrule
Process
& Dialogue turns
& \textbf{4.83$\pm$1.70}
& 5.17$\pm$1.47
& 0.594 \\
\midrule

\multirow[t]{5}{1.45cm}{\raggedright Workload\\(0--20, $\downarrow$)}
& Overall workload
& 11.23$\pm$4.13
& \textbf{7.88$\pm$2.08}
& 0.034 \\
& Mental demand
& 9.25$\pm$5.74
& \textbf{6.00$\pm$3.33}
& 0.036 \\
& Temporal demand
& 12.50$\pm$5.16
& \textbf{7.75$\pm$3.62}
& 0.028 \\
& Effort
& 13.50$\pm$3.48
& \textbf{10.67$\pm$3.42}
& 0.055 \\
& Frustration
& 9.67$\pm$6.14
& \textbf{7.08$\pm$3.50}
& 0.160 \\
\midrule

\multirow[t]{4}{1.45cm}{\raggedright Satisfaction\\(0--20, $\uparrow$)}
& Overall satisfaction
& 11.28$\pm$4.42
& \textbf{14.36$\pm$3.21}
& 0.130 \\
& Self performance
& 12.25$\pm$4.49
& \textbf{14.00$\pm$3.19}
& 0.359 \\
& Tool satisfaction
& 10.58$\pm$5.04
& \textbf{14.58$\pm$3.58}
& 0.079 \\
& Output satisfaction
& 11.00$\pm$4.47
& \textbf{14.50$\pm$3.71}
& 0.121 \\
\midrule
Output quality
& Blind preference rate $\uparrow$
& 35.4\% (17/48)
& \textbf{54.2\% (26/48)}
& -- \\
\bottomrule
\end{tabular}
\end{table}

\subsubsection{Results}

\paragraph{Numerical results.}
% Original: \Cref{tab:e2e-authoring} summarizes the end-to-end results.
As shown in \Cref{tab:e2e-authoring},
\textsc{Ours} significantly reduced NASA-TLX workload relative to \textsc{Baseline} ($7.88{\pm}2.08$ vs.\ $11.23{\pm}4.13$; $p{=}0.034$), while satisfaction ratings also favored \textsc{Ours} but were not statistically significant (11.28$\pm$4.42 vs.\ 14.36$\pm$3.21; $p{=}0.130$).
The number of dialogue turns was similar ($5.17{\pm}1.47$ vs.\ $4.83{\pm}1.70$; $p{=}0.594$), suggesting that the lower workload reflected reduced coordination overhead rather than fewer iterations.
% Original: Blind peer review showed a similar pattern: outputs created with \textsc{Ours} were preferred in 54.2\% of comparisons (26/48), versus 35.4\% (17/48) for \textsc{Baseline}, with 10.4\% (5/48) ties.
% Original: Blind peer review showed that outputs created with \textsc{Ours} were preferred more often than those from \textsc{Baseline} (54.2\% vs.\ 35.4\%), with 10.4\% ties.
In 48 blind pairwise judgments, outputs created with \textsc{Ours} were preferred more often than those from \textsc{Baseline} (54.2\% vs.\ 35.4\%), with 10.4\% ties.
This suggests that the integrated workflow can reduce authoring friction without sacrificing output quality.
Qualitative examples of the authoring results are provided in the Supplementary Section~7.
% Original: Although these quality judgments are descriptive rather than causal, they are consistent with the workload results and suggest that the integrated workflow can reduce authoring friction without sacrificing output quality.
% These descriptive judgments are consistent with the workload results and suggest that the integrated workflow can reduce authoring friction without sacrificing output quality.

\paragraph{Qualitative feedback.}
Participants mainly described \textsc{Ours} as easier to work with because retrieval, preview, and iterative refinement remained in a single workspace.
The most common positive comment was a \emph{``one-stop''} experience (3 participants), followed by novice-friendly exemplar guidance (2) and better support for polishing simple aesthetics (2).
In contrast, the dominant complaint about \textsc{Baseline} was \emph{multi-page operation and manual overhead} (4 participants), along with mismatches between retrieved exemplars and generated SVGs (2), and limited conversational memory (2).
\Cref{fig:system-ui} illustrates this contrast: in \textsc{Ours}, participants could keep multiple exemplars visible while iteratively inspecting and refining SVG drafts in place, whereas the baseline required manually moving artifacts across the retriever, chat interface, and external viewer.
At the same time, several sessions revealed a \emph{``debugging over designing''} pattern: later iterations often focused on repairing conversion issues such as missing pictograms, overwritten labels, or misaligned layouts rather than exploring richer alternatives.
This suggests that current image-to-SVG generation already supports reasonably faithful exemplar adaptation, but higher-level creative exploration and aesthetic enrichment remain limited.

\section{Discussion and Limitations}
\label{sec:discussion}

Taken together, the retrieval experiments demonstrate that our method improves retrieval quality, and the authoring study suggests that these gains translate into more effective infographic authoring.
Meanwhile, the results point to some limitations and opportunities for future work.\looseness=-1

\paragraph{Adaptation fidelity.}
While the integrated workflow supports exemplar reuse, adaptation quality still depends heavily on the robustness of the downstream SVG editing pipeline.
As observed in the authoring study, users sometimes shifted from designing to debugging, repairing issues such as missing icons, overwritten labels, misaligned layouts, or disproportionate marks.
These failures suggest that the current adaptation pipeline is still fragile at the execution level.
Even when the intended direction is clear, the system does not always apply edits faithfully and cleanly.
As P9 commented, ``Once I know what I want to change, I still need the system to execute that change more reliably. Right now, some of the work becomes checking whether the generated SVG broke something else.''
This suggests the need for more robust structural representations, stronger edit planning, and constraint-aware SVG editing to improve the reliability of exemplar-based adaptation.
% Better structural representations, edit planning, and constraint-aware SVG editing are needed to make exemplar reuse more reliable.

\paragraph{Draft quality and downstream handoff.}
Beyond execution-level errors, the system is better positioned as a tool for ideation and first-draft creation than as a substitute for professional refinement.
P2 remarked, ``It gets me to a first draft very quickly, which is already valuable, but I would still want to polish the spacing, alignment, and typography myself before using it in a final design.''
P7 similarly observed, ``This feels strongest when I am still exploring options. Once I know what I want, I start wanting much finer control over individual elements than the current workflow gives me.''
Together, these comments suggest that the main value of the current system lies in accelerating exploration, reference combination, and early draft construction, rather than in producing fully polished artifacts without further intervention.
Accordingly, rather than treating SVG generation as the endpoint of the workflow, future work should support a more seamless handoff from generated drafts to professional editing environments, where designers can refine typography, spacing, alignment, and other details.

\section{Conclusion}
\label{sec:conclusion}

We presented an intent-aware infographic retrieval framework for design inspiration and exemplar-based authoring.
Motivated by a formative study of how users describe desired infographics, our method represents queries with multiple intent facets, rewrites and weights these facets explicitly, and performs facet-aware matching over an infographic corpus.
We further integrated retrieval with SVG-based adaptation in a conversational workflow that supports search, selection, reuse, and iterative refinement. % in one interface
The evaluation results demonstrate our method consistently outperformed strong retrieval baselines and reduced authoring workload while maintaining or improving output quality.
We hope this work helps move infographic authoring support from generic search toward intent-aware, exemplar-centered systems that better align with how people actually seek and reuse visual inspiration.\looseness=-1

%% if specified like this the section will be omitted in review mode
% \acknowledgments{%
% 	The authors wish to thank A, B, and C.
%   This work was supported in part by a grant from XYZ (\# 12345-67890).%
% }

\section*{SUPPLEMENTAL MATERIALS}
The supplementary document includes
(1) the query-parsing prompt templates and structured five-facet schema,
(2) the chart-type soft-matching table and taxonomy overview,
(3) additional details of the interactive interface and the SVG handling mechanism,
(4) the human-written query and human-judgment protocols used in the evaluation, and
(5) selected authoring-system case pairs discussed in the paper.
The code will also be included in the supplementary package.

\bibliographystyle{abbrv-doi-hyperref}
\bibliography{template}

% \appendix % You can use the `hideappendix` class option to skip everything after \appendix

% \section{About Appendices}
% Refer to \cref{sec:appendices_inst} for instructions regarding appendices.

% \section{Troubleshooting}
% \label{appendix:troubleshooting}

% \subsection{ifpdf error}

% If you receive compilation errors along the lines of \texttt{Package ifpdf Error: Name clash, \textbackslash ifpdf is already defined} then please add a new line \verb|\let\ifpdf\relax| right after the \verb|\documentclass[journal]{vgtc}| call.
% Note that your error is due to packages you use that define \verb|\ifpdf| which is obsolete (the result is that \verb|\ifpdf| is defined twice); these packages should be changed to use \verb|ifpdf| package instead.

% \subsection{\texttt{pdfendlink} error}

% Occasionally (for some \LaTeX\ distributions) this hyper-linked bib\TeX\ style may lead to \textbf{compilation errors} (\texttt{pdfendlink ended up in different nesting level ...}) if a reference entry is broken across two pages (due to a bug in \verb|hyperref|).
% In this case, make sure you have the latest version of the \verb|hyperref| package (i.e.\ update your \LaTeX\ installation/packages) or, alternatively, revert back to \verb|\bibliographystyle{abbrv-doi}| (at the expense of removing hyperlinks from the bibliography) and try \verb|\bibliographystyle{abbrv-doi-hyperref}| again after some more editing.

\end{document}